\input{seam.sty}
\def\beq{\begin{equation}}
\def\eeq{\end{equation}}

\def\addots{\mathinner{\mkern1mu\raise1pt\vbox{\kern7pt\hbox{.}}\mkern2mu
\raise4pt\hbox{.}\mkern2mu\raise7pt\hbox{.}\mkern1mu}}

\def\Exp{\mbox{Exp}}

\def\upsilon{\mu}
\def\varsigma{\nu}

\numberwithin{equation}{section}

\begin{document}
\begin{center}
\title{Lattice Realizations of the Open Descendants\\ 
of Twisted Boundary Conditions\\
for $sl(2)$ $A$-$D$-$E$ Models
}
\author{C.H. Otto Chui\footnote{Email: C.Chui@ictp.trieste.it}
and Paul A. Pearce\footnote{Email: P.Pearce@ms.unimelb.edu.au}} 
\address{ ${}^1$
The Abdus Salam International Center for Theoretical Physics, \\
Strada Costiera 11, Trieste 34014, Italy}
\address{${}^2$
Department of Mathematics and Statistics, University of Melbourne\\ 
Parkville, Victoria 3010, Australia}
\end{center}
\begin{abstract}
The twisted boundary conditions and associated partition functions of the conformal $sl(2)$ \ade models are studied on the Klein bottle and the M\"obius strip. 
The \ade minimal lattice models give realization to the complete classification of the open descendants of the $sl(2)$ minimal theories. We construct the transfer matrices of these lattice models that are consistent with non-orientable geometries.  In particular, we show that in order to realize all the Klein bottle amplitudes of different crosscap states, not only the topological flip on the lattice but also the involution in the spin configuration space must be taken into account. This involution is the $\Bbb Z_2$ symmetry of the Dynkin diagrams which corresponds to the simple current of the Ocneanu algebra.  
\end{abstract}

\section{Introduction}

The study of conformal field theories on non-orientable surfaces has attracted much 
attention since the discovery of D-branes~\cite{Polchinski95}.
Open descendants~\cite{Sagnotti88} is a systematic construction of conformal field theories on  
non-orientable surfaces from the oriented ones by 
requiring the theories to satisfy certain consistency conditions~\cite{SS03}.
The open descendants of many different models have been studied~\cite{PSS95a,PSS95,SS97,HSS02,GM04}.
Moreover, it was discovered that new crosscap states can be found by the approach of 
simple current invariance~\cite{HSS99,FHSSW00,HS00}. 
Conformal defects on non-orientable surfaces were discussed in \cite{FRS04}.

On the other hand, it is well known that statistical mechanics models realize conformal
field theories in the continuum scaling limit. In particular, the lattice realizations~\cite{CMOP,CMP03} of conformal \ade unitary minimal models on a torus with
defect lines~\cite{PZ0011021, PZ01}  has been studied. The Ising model on the Klein bottle and
M\"obius strip has also been investigated~\cite{CP02,LuWu}. 
In this article, we study the critical $sl(2)$ \ade minimal lattice models on the Klein bottle
and on the M\"obius strip with twisted boundary conditions. 
All the lattice realizations of the known open descendants for the $sl(2)$ models
are obtained in these non-orientable geometries.

The paper is organized as follows. In section \ref{sec:ConformalBC}, we briefly review the results of
the boundaries conditions of the $sl(2)$ models on a torus (section \ref{sec:Torus}) and on a cylinder (section \ref{sec:Cylinder}).
In section \ref{sec:OpenDescendants}, we review the open descendants on the Klein bottle and the M\"obius strip.
In section \ref{sec:OpenDescendantsDerivation}, we describe in detail how to compute the crosscap coefficients and to obtain the partition functions on the Klein bottle and the M\"obius strip.
In section \ref{sec:LatticeModels}, we define the \ade lattice models and construct the transfer matrices
on the Klein bottle and the M\"obius strip. 
In section \ref{sec:Computation}, we summarize our numerical results.

\section{Conformal Boundary Conditions of $sl(2)$ Models} \label{sec:ConformalBC}

The chiral algebra of an $SU(2)_k$ Wess-Zumino-Witten model
is given by the affine algebra $\hat{sl}(2)$ at level $k$ and there exists
a finite set $\mathcal I=\{1,2,\ldots,k+1\}$ of irreducible representations
$\mathcal V_i$, $i\in \mathcal I$.
The central charge is $c=3k/(k+2)$ and the conformal weights are
$h_j=(j^2-1)/4(k+2)$, $j\in\mathcal I$.

It is well known that the modular invariant partition functions of the 
$SU(2)_k$ WZW models are classified~\cite{CIZ87a} according to the Dynkin diagrams 
of the classical \ade simply laced 
Lie algebras, where the Coxeter number $g$ is identified with $k+2$.

\begin{figure}[htb]
\begin{center}
    \setlength{\unitlength}{.5cm}
\begin{equation*}
\mbox{}\hspace{1.1in}\mbox{}
\begin{array}{ccccc}
    \text{\text{Graph $G$\qquad}}& \mbox{}\quad\qquad\text{$g$}\quad\qquad\mbox{}&\text{$\Exp(G)$}&\text{Type/$H$}&\Gamma\\[10pt]
   \begin{picture}(6,1)
        \put(-3.5,0){\p{}{A_{L}}}
        \put(0,0){\line(1,0){4}}
        \multiput(1,0)(1,0){2}{\pp{}{\bullet}}
        \put(4,0){\pp{}{\bullet}}
        \put(0,0){\pp{}{*}}
        \put(1,0){\pp{}{\square}}
       \put(0,.3){\pp{b}{1}}
        \put(1,.3){\pp{b}{2}}
        \put(2,.3){\pp{b}{3}}
        \put(3,.3){\pp{b}{\cdots}}
        \put(4,.3){\pp{b}{L}}
    \end{picture}  & L+1 & 1, 2, \cdots, L &\mbox{I}&{\Bbb Z_2}\\
     \begin{picture}(6,2)
        \put(-3.5,0){\p{}{D_{\ell +2}\,\mbox{($\ell$ even)}}}
        \put(0,0){\line(1,0){3.5}}
        \put(3.5,0){\line(1,1){1}}
        \put(3.5,0){\line(1,-1){1}}
        \multiput(1,0)(1,0){2}{\pp{}{\bullet}}
        \put(3.5,0){\pp{}{\bullet}}
        \put(4.5,1){\pp{}{\bullet}}
        \put(4.5,-1){\pp{}{\bullet}}
        \put(-.2,-.04){\pp{l}{*}}
        \put(.75,-.04){\pp{l}{\square}}
        \put(0,.3){\pp{b}{1}}
        \put(1,.3){\pp{b}{2}}
        \put(2,.3){\pp{b}{3}}
        \put(2.75,.3){\pp{b}{\cdots}}
        \put(3.5,.3){\pp{b}{\ell}}
        \put(4.5,1){\pp{l}{~\ell+1}}
        \put(4.5,-.7){\pp{l}{~\ell+2}}
    \end{picture}    & 2\ell +2 & 1, 3, \cdots, 2\ell +1, \ell +1 &\mbox{I}&{\Bbb Z_2} \\[10pt]
 \begin{picture}(6,2)
        \put(-3.5,0){\p{}{D_{\ell +2}\,\mbox{($\ell$ odd)}}}
        \put(0,0){\line(1,0){3.5}}
        \put(3.5,0){\line(1,1){1}}
        \put(3.5,0){\line(1,-1){1}}
        \multiput(0,0)(1,0){3}{\pp{}{\bullet}}
        \put(3.5,0){\pp{}{\bullet}}
        \put(4.5,1){\pp{}{\bullet}}
        \put(0,.3){\pp{b}{1}}
        \put(1,.3){\pp{b}{2}}
        \put(2,.3){\pp{b}{3}}
        \put(2.75,.3){\pp{b}{\cdots}}
        \put(3.5,.3){\pp{b}{\ell}}
        \put(4.5,1){\pp{l}{~\ell+1}}
        \put(4.5,-.7){\pp{l}{~\ell+2}}
        \put(4.35,-1.05){\pp{l}{*}}
        \put(3.25,-.05){\pp{l}{\square}}
    \end{picture}    & 2\ell +2 & 1, 3, \cdots, 2\ell +1, \ell +1 &\mbox{II}/A_{2\ell+1} &{\Bbb Z_2}\\
   \begin{picture}(6,2.5)
        \put(-3.5,0){\p{}{E_{6}}}
        \put(0,0){\line(1,0){4}}
        \put(2,0){\line(0,1){1}}
        \multiput(1,0)(1,0){4}{\pp{}{\bullet}}
        \put(2,1){\pp{}{\bullet}}
        \put(0,0){\pp{}{*}}
        \put(1,0){\pp{}{\square}}
        \put(0,.3){\pp{b}{1}}
        \put(1,.3){\pp{b}{2}}
        \put(2,-.3){\pp{t}{3}}
        \put(3,.3){\pp{b}{4}}
        \put(4,.3){\pp{b}{5}}
        \put(2,1.3){\pp{b}{6}}
     \end{picture}  & 12 & 1, 4, 5, 7, 8, 11 &\mbox{I}&{\Bbb Z_2} \\
   \begin{picture}(6,2.5)
        \put(-3.5,0){\p{}{E_{7}}}
        \put(0,0){\line(1,0){5}}
        \put(3,0){\line(0,1){1}}
        \multiput(1,0)(1,0){5}{\pp{}{\bullet}}
         \put(3,1){\pp{}{\bullet}}
        \put(0,.3){\pp{b}{1}}
        \put(1,.3){\pp{b}{2}}
        \put(2,.3){\pp{b}{3}}
        \put(3,-.3){\pp{t}{4}}
        \put(4,.3){\pp{b}{5}}
        \put(5,.3){\pp{b}{6}}
        \put(3,1.3){\pp{b}{7}}
        \put(0,0){\pp{}{*}}
        \put(1,0){\pp{}{\square}}
     \end{picture}  & 18 & 1, 5, 7, 9, 11, 13, 17  &\mbox{II}/D_{10}&1\\
   \begin{picture}(6,2.5)
        \put(-3.5,0){\p{}{E_{8}}}
        \put(0,0){\line(1,0){6}}
        \put(4,0){\line(0,1){1}}
        \multiput(1,0)(1,0){6}{\pp{}{\bullet}}
        \put(4,1){\pp{}{\bullet}}
        \put(0,0){\pp{}{*}}
        \put(1,0){\pp{}{\square}}
        \put(0,.3){\pp{b}{1}}
        \put(1,.3){\pp{b}{2}}
        \put(2,.3){\pp{b}{3}}
        \put(3,.3){\pp{b}{4}}
        \put(4,-.3){\pp{t}{5}}
        \put(5,.3){\pp{b}{6}}
        \put(6,.3){\pp{b}{7}}
        \put(4,1.3){\pp{b}{8}}
     \end{picture}  & 30 & 1, 7, 11, 13, 17, 19, 23, 39&\mbox{I}&1\\
\end{array}
\end{equation*}
\end{center}
\caption{\ade graphs corresponding to the Dynkin diagrams of the classical \ade simply laced 
Lie algebras. The nodes associated with the identity and the fundamental are 
shown by $*$, $\square$ respectively. Also shown are the Coxeter numbers $g$, exponents $\Exp(G)$, the Type I or II, the parent graphs $H\ne G$ and the diagram automorphism group $\Gamma$. The $D_4$ graph is exceptional having the automorphism group $\Bbb S_3$.}
\label{tbl:Graphs}
\end{figure}

On an infinitely long cylinder, the Hilbert space $\mathcal H$ can be expressed as a 
finite sum of the irreducible representations of the tensor product
of the left and the right chiral algebra
\beq
\mathcal H=\bigoplus Z_{j\,\bar j} \mathcal V_j\otimes \bar{\mathcal V}_{\bar j}
\eeq
where $Z_{j\,\bar j}$ are non-negative integer multiplicities. The diagonal subset of the labels $(j,\bar j)$
of $Z_{j\,\bar j}$
\beq
\mathcal E=\{(j,\alpha) |\alpha=1,\ldots,Z_{j\,j}\} 
\eeq
is given by the exponents of the Dynkin diagram $G$ of \ade type, where
$\alpha$ is the multiplicity index.

\subsection{Twisted Boundary Conditions on a Torus} \label{sec:Torus}

In this section, we briefly recall the results on the twisted boundary conditions.
Further details are given in \cite{PZ0011021,PZ01,CMP03,CoqTwist01}.

For a WZW model associated with the Dynkin diagram $G$,
the twisted boundary states $x$ on the torus are labeled by the nodes of the 
Ocneanu graph $x\in\tilde G$ \cite{Ocn2}. The corresponding twisted operators $X_x$ commute
with the Virasoro algebra
\beq
[L_n,X_x]=[{\bar L}_n,X_x]=0
\eeq

The Hilbert space with a twisted boundary $x$ inserted is given by
\beq
\mathcal H_{x} = \bigoplus_{i,\bar i\in\mathcal I} \tilde V_{i,\bar i^*;\,x}{}^1\,
\mathcal V_i \otimes \bar{\mathcal V}_{\bar i}
\label{eq:Hilbertspace}
\eeq
where $\tilde V_{i,\bar i^*;\,x}{}^1$ are non-negative integers. 
The twisted partition functions are labeled by 
\begin{equation}
x=(a,b,\kappa)\in(H,\bar H,\Bbb Z_2)
\end{equation}
where $H$ ($\bar H$) is  the left (right) chiral copy of the parent graph of $G$ and $\kappa=1,2$ labels 
the $\Bbb Z_2$ automorphisms of $G$
(figure \ref{tbl:Graphs}). 

The twisted partition functions in the direct channel are given by the toric matrices
$P_{ab}^{(\kappa)}$
\begin{equation}
Z_{(a,b,\kappa)}(q)=\sum_{i,\bar i}
[P_{ab}^{(\kappa)}]_{i \bar i}\;
\chi_{i}(q)\,\chi_{\bar i}(\overline{q}),\quad
\label{PZtwistedPFs}
\end{equation}
\begin{equation}
[P_{ab}^{(\kappa)}]_{i \bar i}=\sum_{c\in T_\kappa} n_{i a}^{(H)}{}^c\,n_{\bar i
b}^{(H)}{}^{\zeta(c)}
\label{toricmatrix}
\end{equation}
and $[P_{ab}^{(\kappa)}]_{i\bar i}=\tilde V_{i,\bar i^*;\,x}{}^1$. 
Let us explain the notations in 
\eqref{toricmatrix}.
$n_i^{(H)}=(n_{i a}^{(H)}{}^b)$ are the fused adjacency matrices of the Type I parent graph $H$ of $G$.
The fused adjacency matrices $n_i\equiv n_i^{(G)}$ of a graph $G$
are defined by the $sl(2)$ recurrence relation
\begin{equation}
    n_{1}=\text{I},\qquad n_{2}=G,\qquad n_{i+1}=n_{2}\,n_{i}-n_{i-1}\,
    \ \text{ for }\  2<i< g-1,
    \label{eq:ni}
\end{equation}
where $n_2=G$ is the adjacency matrix of the graph $G$.
$T_{\kappa}$ are the following subsets of $A_{g-1}$ 
\begin{eqnarray}
T_1&=&\cases{\{1,2,\ldots,L\},&G=A_L\cr
\{1,3,5,\ldots,2\ell-1,2\ell\},&G=D_{2\ell}\cr
\{1,2,3,\ldots,4\ell-1\},&G=D_{2\ell+1}\cr
\{1,5,6\},&G=E_6\cr
\{1,3,5,7,9,10\},&G=E_7\cr
\{1,7\},&G=E_8}
\end{eqnarray}
and 
\begin{eqnarray}
T_2&=&\cases{
\{2,4,\ldots,2\ell-2\},&G=D_{2\ell}\cr
\;T_1,&\mbox{otherwise}}
\end{eqnarray}
The involutive twist $\zeta$ is the identity for Type I theories but for Type II theories has the action
\begin{equation}
\zeta=\cases{s\mapsto 4\ell-s,\ s=2,4,\ldots,2\ell-2,&G=D_{2\ell+1}\cr
\{1,3,5,7,9,10\}\mapsto \{1,9,5,7,3,10\},&G=E_7
}
\end{equation}
In particular, $Z_{(1,1,1)}$ is the modular invariant partition function with no twist.

\subsection{Cylinder Boundary Conditions} \label{sec:Cylinder}

The complete set of conformal boundary conditions $|a\!>$ for $sl(2)$ models 
on the cylinder are labeled by
the nodes of the graph 
$a\in G$ \cite{BPPZ00}. They are linear combinations of the Ishibashi states $|j\!\gg_B$
\cite{Ishibashi89}
\beq
|a\!>=\sum_{j\in\mathcal E}B_a^j\, |j\!\gg_B\, ,\qquad
B_a^j=\frac{\psi_a^j}{\sqrt{S_1^j}}
\label{eq:boundaryIshibashi}
\eeq 
where $S_i^j=\sqrt{\frac{2}{g}}\sin \frac{\pi ij}{g}$ is the modular matrix and 
the expressions of the coefficients $\psi_a^j$ can be found in the appendix of \cite{BPPZ00}. 

The partition functions in the transverse channel are given by
\beq
A_{b|a}(\tilde q)=\ <\!b|{\tilde q}^{\frac{1}{2}(L_0+\bar{L}_0-\frac{c}{12})}|a\!>\ =\ \sum_{j\in\mathcal E} B_a^j (B_b^j)^ *  \chi_{j}(\tilde q)
\label{eq:transverseAnnulus}
\eeq
and in the direct channel
\beq
A_{b|a}(q)=\sum_{i\in\mathcal I} n_{i\,a}{}^b \chi_i(q)
\label{eq:directAnnulus}
\eeq
where $n_{i\,a}{}^b$ are non-negative integers given by \eqref{eq:ni}.
The direct and the transverse channels are related by the modular transformation $S$. Comparing
\eqref{eq:directAnnulus} with \eqref{eq:transverseAnnulus} shows that
$n_{i\,a}{}^b$ satisfy the Verlinde-like formula: 
\beq 
n_{i\,a}{}^b= \sum_{j\in\mathcal E} S_i^j B_a^j (B_b^{j})^* = 
\sum_{j\in\mathcal E} \frac{S_i^j}{S_1^j} \psi_a^j (\psi_b^{j})^*
\label{eq:Verlinde-like}
\eeq

For a cylinder with boundaries $a$ and $b$ and a twisted operator $x$ inserted,
the twisted partition function is \cite{PZ0011021}
\beq
A^{(x)}_{b|a}(q)=\sum_{i\in\mathcal I} ({\tilde n}_x \cdot n_i)_{a}{}^b \chi_i(q)
\label{eq:simplecurrentAnnulus}
\eeq
where $x\in\tilde G$ and ${\tilde n}_x$ is the corresponding $|G|\times|G|$ non-negative integer
matrix representation. The formulae of ${\tilde n}_x$ are derived in \cite{PZ00a} and their
explicit expressions are shown in \cite{CMP03}.

Note that $A^{(x)}_{b|a}(q)$ can be expressed as
a linear combination of the non-twisted partition functions \eqref{eq:directAnnulus} since
the twisted operator can propagate to the boundary to produce an integrable boundary condition
\cite{CMP03}.

\section{Open Descendants}\label{sec:OpenDescendants}

The complete classification for the open descendants of the 
modular invariant partition functions for the 
$sl(2)$ models 
has been carried out by Pradisi, Sagnotti and Stanev \cite{PSS95a,PSS95}.
In this section, we consider the open descendants 
of twisted boundary conditions.

For an $sl(2)$ model, the crosscap states $|\kappa\!>$
are labeled by the $\Bbb Z_2$ automorphisms of the graph $G$,
with $\kappa=1,2$. These $\Bbb Z_2$ automorphisms correspond to
the simple currents \cite{SY00} of the Ocneanu algebra $\tilde G$. Similar to the boundary
states \eqref{eq:boundaryIshibashi}, the crosscap states are
linear combinations of the Ishibashi states $|j\!\gg_C$
\beq
|\kappa\!>=\sum_{j\in\mathcal E} \Gamma^{(\kappa)}_j |j\!\gg_C
\eeq 
where the crosscap coefficients  $\Gamma^{(\kappa)}_j$ 
will be determined in section \ref{sec:OpenDescendantsDerivation}.
Note that, unlike the case of the boundary states $|a\!>$ in which different boundary states can be present simultaneously in a conformal field theory, only one
type of crosscap state is allowed~\cite{FPS94} due to crosscap constraints.

\subsection{Klein bottle}\label{sec:OD:KleinBottle}

To construct a closed non-oriented CFT on the Klein bottle from 
the closed oriented theory on the torus, let us start from  \eqref{eq:Hilbertspace}.
On the Klein bottle, due to the non-orientability, 
the left chiral fields exchange with the right chiral fields. Essentially
all the non-diagonal fields 
in the Hilbert space \eqref{eq:Hilbertspace} are projected out, and only the diagonal terms survive.
Furthermore, for a CFT on the Klein bottle to be consistent with the non-orientable topology,
the corresponding twisted partition function on the torus must be symmetric, i.\,e.\ 
$\tilde V_{i, j;\,x}{}^1=\tilde V_{j,i;\,x}{}^1$.

On the Klein bottle, the twisted boundary conditions are labeled by $a\in H$, where
$H$ is the parent graph of $G$. It suffices to consider the case of two twisted operators inserted. 
In the transverse channel, the Klein bottle is viewed as a cylinder with two crosscaps at the ends. The 
transverse channel twisted Klein bottle partition function with
twisted operators $a$ and $b$ inserted is given by
\beq
K^{(\kappa)}_{(a,b)}(\tilde q) =\sum_{j\in\mathcal E} \Gamma_j^{(\kappa)} \Gamma_j^{(\kappa)*}
\frac{\psi_a^{(H) j}}{\psi_1^{(H) j}}
\left(\frac{\psi_b^{(H) j}}{\psi_1^{(H) j}}\right)^*
\chi_j(\tilde q)
\eeq
where $\psi_a^{(H) j}$ are the boundary coefficients of the parent
graph $H$ appeared in \eqref{eq:boundaryIshibashi}.
On the other hand, in the direct channel, the twisted partition function is 
\beq
K^{(\kappa)}_{(a,b)}(q)=\sum_{i\in\mathcal I} K^{(\kappa)}_{ia}{}^b \chi_i(q)
\eeq
where $K^{(\kappa)}_{ia}{}^b$ are the integer Klein bottle coefficients.
The two channels are related by the modular transformation $S$ \cite{Sagnotti88}, so that 
$K^{(\kappa)}_{ia}{}^b$ equals
\beq
K^{(\kappa)}_{ia}{}^b =\sum_{j\in\mathcal E} S_i^j\Gamma_j^{(\kappa)} \Gamma_j^{(\kappa)*}
\frac{\psi_a^{(H) j}}{\psi_1^{(H) j}}
\left(\frac{\psi_b^{(H) j}}{\psi_1^{(H) j}}\right)^*
\eeq

The Klein bottle coefficients  $K^{(\kappa)}_{ia}{}^b$ satisfy the following property
\beq
K^{(\kappa)}_{ia}{}^b=K^{(\kappa)}_{ib}{}^a=K^{(\kappa)}_{i\,ab}{}^1=
\sum_{c\in H} {\hat N}_{ac}^{(H)}{}^b K^{(\kappa)}_{ic}{}^1
\eeq
where ${\hat N}_a^{(H)}=({\hat N}_{ab}^{(H)}{}^c)$ are the graph fusion matrices
of the parent graph $H$ given by the Verlinde-like formula
\beq
{\hat N}_{ab}{}^c=\sum_{j\in\mathcal E} \frac{\psi_a^j\psi_b^j(\psi_c^j)^*}{\psi_*^j}
\eeq
where $*$ denotes the fundamental node of the graph (figure \ref{tbl:Graphs}).
In general, the graph fusion matrices of a graph $G$ form the graph fusion algebra of $G$
\beq
{\hat N}_{a} {\hat N}_{b}= \sum_{c\in G} {\hat N}_{ab}{}^c {\hat N}_{c}
\eeq
In particular, for a Type I graph $H$, its graph fusion matrices are of non-negative
integer value.


For simplicity, we consider a single twisted operator inserted on the Klein bottle.
The direct channel twisted partition 
$K^{(\kappa)}_{(a,1)}(q)$ is projected~\footnote{The exception is $K^{(2)}_{(2\ell-1,1)}=K^{(2)}_{(2\ell,1)}$ for $D_{2\ell}$
whose torus counterpart is $Z_{(2\ell-1,2\ell,1)}$. We will discuss this
from the lattice point of view in section \ref{sec:KleinD2l}.} from the torus counterpart
$Z_{(a,a,1)}(q)$.
The complete non-oriented closed string partition function is half of the sum of the 
Klein bottle and the torus partition functions in the direct channel 
\beq
Z^{\mbox{\scriptsize{non-oriented}}}_{(a,\kappa)}=\frac{1}{2}(Z^{\mbox{\scriptsize{torus}}}_{(a,a,1)} + 
K^{(\kappa)}_{(a,1)})
\label{eq:unorientedpartitionfunction}
\eeq
with non-negative integer coefficients. Hence, together with the fact that the Klein bottle
coefficients are projected from the torus counterparts, they must satisfy the following 
constraints
\beq
|K^{(\kappa)}_{ia}{}^1| \leq [P_{aa}^{(1)}]_{ii}
\qquad\mbox{and}\qquad
K^{(\kappa)}_{ia}{}^1\equiv [P_{aa}^{(1)}]_{ii} \,\mod 2
\label{eq:Kleinbottleconstraint}
\eeq

\subsection{M\"obius strip}\label{sec:OD:Moebius}

On the M\"obius strip, one works with the ``tilde'' character $\tilde\chi(q)$.
Recall that the character $\chi(q)$ can be expressed as a polynomial with an overall phase factor
\beq
\chi_i(q)=q^{h_i-c/24} \sum_k d_i(k) q^k
\eeq
The ``tilde'' character $\tilde\chi(q)$
is defined by mapping $q\rightarrow e^{i\pi}q$ but with the overall phase factor unchanged
\beq
\tilde \chi_i(q)=q^{h_i-c/24} \sum_k (-1)^k d_i(k) q^k
\eeq

In the transverse channel, the M\"obius strip is viewed as a cylinder with
a crosscap and a boundary at the ends. The  transverse channel partition function
is given by
\beq
M^{(\kappa)}_{a}(\tilde q) = \sum_{j\in\mathcal E} B_a^j  \Gamma_j^{(\kappa)} {\tilde\chi}_j(\tilde q)
\eeq
whereas the direct channel partition function is 
\beq
M^{(\kappa)}_{a}(q) = \sum_{i\in\mathcal I} M^{(\kappa)}_{i a} {\tilde\chi}_j(q)
\eeq
where $M^{(\kappa)}_{i a}$ are integer M\"obius strip coefficients.
The two channels are transformed by the modular matrix $P$ \cite{Sagnotti88,FRS04}
\beq
P=\sqrt{T}ST^2S\sqrt{T}
\label{eq:Pmatrixgeneral}
\eeq
which satisfies $P^2=C$, where $C$ is the conjugation matrix $C_{ij}=\delta_{j,i^*}$.
In \eqref{eq:Pmatrixgeneral},
$\sqrt{T}$ is defined as $(\sqrt{T})_{ij}=e^{i\pi (h_i-c/24)} \delta_{i,j}$.

Then the M\"obius strip coefficients equal
\beq
M^{(\kappa)}_{i a}= \sum_{j\in\mathcal E} P_i^j B_a^j \Gamma_j^{(\kappa)} 
\label{eq:Moebiusstripcoefficient}
\eeq
%
%
The direct channel partition function $M^{(\kappa)}_{a}(q)$ is projected from the cylinder counterpart
$A_{a|a}^{(\kappa)}(q)$, where $\kappa=1$ is an identity and $\kappa=2$ denotes the simple current
of $\tilde G$ which corresponds to the $\Bbb Z_2$ symmetry of the graph $G$. Explicitly,
\beq
A^{(\kappa)}_{b|a}(q)=\sum_{i\in\mathcal I} (\sigma^{\kappa-1} \cdot n_i)_{a}{}^b \chi_i(q)
\eeq
where $\sigma$ is the order 2 permutation matrix which encodes the $\Bbb Z_2$ automorphism. 
For convenience, we use the notation $\sigma(a)$ to denote the action of the $\Bbb Z_2$ automorphism
on the node $a$ of $G$. 

The complete non-oriented open string partition function is half of the sum
of the cylinder and the M\"obius strip partition functions 
\beq
Z^{\mbox{\scriptsize{open}}\,(\kappa)}_a=\frac{1}{2}(A^{(\kappa)}_{a|a}\pm M_{a}^{(\kappa)})
\label{eq:openunorientedpartitionfunction}
\eeq
where the sign of $M_{a}^{(\kappa)}$ cannot be determined from CFT \cite{Stanev03}.
Thus the M\"obius strip coefficients are subject to the following constraints:
\begin{equation}
|M^{(\kappa)}_{i a}|\leq (\sigma^{\kappa-1} \cdot n_i)_{a}{}^a \qquad\mbox{and}\qquad
M^{(\kappa)}_{i a}\equiv (\sigma^{\kappa-1} \cdot n_i)_{a}{}^a \,\mod 2
\label{eq:Moebiusstripconstraint}
\end{equation}

\subsection{Derivation of the crosscap coefficients from M\"obius strip}\label{sec:OpenDescendantsDerivation}

In this section, we follow the method 
in \cite{HS00} to compute the crosscap coefficients.
The idea of the method is to find solutions to the M\"obius strip with vacuum boundary that 
are consistent with the integrality and positivity conditions of the 
Klein bottle \eqref{eq:Kleinbottleconstraint} and  of the M\"obus strip \eqref{eq:Moebiusstripconstraint}. Once the M\"obius strip coefficients with vacuum boundary 
are obtained, one can take  an inverse of $P$ 
in \eqref{eq:Moebiusstripcoefficient} to get the crosscap coefficients.

In general,
for a M\"obius strip with boundary state $a$, we multiply both sides of \eqref{eq:Moebiusstripcoefficient}
by $P$ to obtain  the expressions for $\Gamma_k^{(\kappa)}$
\beq
\sum_{i\in\mathcal I} P_k^i M_{i a}^{(\kappa)}
=\sum_{j\in \mathcal E} \delta_{k,j} B_a^j\Gamma_j^{(\kappa)}
= \sum_{\alpha} B_a^{(k,\alpha)}\Gamma_{(k,\alpha)}^{(\kappa)}
\label{eq:crosscapequationraw}
\eeq
where $\alpha$ are the multiplicities of the exponents $k\in\mathcal E$. 
Note that for $sl(2)$ WZW models, $\alpha=1$ except for the case of $k=g/2$ for $D_{2\ell}$ which
is of degeneracy 2. 
Shortly we will see that  $\Gamma_k^{(\kappa)}=0$ for $k\notin\mathcal E$.

For a cylinder with vacuum boundary state, the partition function is 
\beq
A^{(\kappa)}_{1|1}(q)=\sum_{i\in\mathcal I} (\sigma^{\kappa-1} \cdot n_i)_{1}{}^1 \chi_i(q)
\eeq
The vacuum boundary is non-degenerate so that
$(\sigma^{\kappa-1}\cdot n_i)_1{}^1=0,1$. Positivity of the M\"obius strip \eqref{eq:Moebiusstripconstraint}
requires
that the M\"obius strip coefficients with vacuum boundary state must satisfy
\beq
M_{i 1}^{(\kappa)}=\eta_i^{(\kappa)}  (\sigma^{\kappa-1} \cdot n_i)_{1}{}^1\, ,\qquad \eta_i^{(\kappa)}=\pm1
\eeq
Thus,
\beq
\Gamma_k^{(\kappa)}= \frac{1}{B_1^k} \sum_{i\in\mathcal I} P_k^i M_{i1}^{(\kappa)}
=\frac{\sqrt{S_1^k}}{\psi_1^k} \sum_{i\in\mathcal I} \eta_i^{(\kappa)}\, 
(\sigma^{\kappa-1} \cdot n_i)_{1}{}^1 P_k^i
\, ,\qquad k\neq\frac{g}{2} \mbox{ when } G=D_{2\ell}
\label{eq:Gammafromvacuum}
\eeq
Later, we will show that for $k=\frac{g}{2}$ when  $G=D_{2\ell}$
\beq
\Gamma_{\frac{g}{2}^+}^{(\kappa)}=\Gamma_{\frac{g}{2}^-}^{(\kappa)}
=\frac{1}{2}\frac{\sqrt{S_1^{\frac{g}{2}}} }{\psi_1^{\frac{g}{2}}}
 \sum_{i\in\mathcal I} \eta_i^{(\kappa)}\, 
(\sigma^{\kappa-1} \cdot n_i)_{1}{}^1 P_{\frac{g}{2}}^i\, ,
\qquad G=D_{2\ell}
\eeq

In the following section, we will determine the signs of $\eta_{i}^{(\kappa)}$ 
up to an overall sign of $\Gamma_k^{(\kappa)}$
by 
positivity of the Klein bottle coefficients \eqref{eq:Kleinbottleconstraint}
and consistency with the solutions of $\Gamma_k^{(\kappa)}$ obtaining from
boundary states other than the vacuum.

Before we proceed, we list the properties of $P$ that will be used later on. 
For $s\ell(2)$ WZW models,
$P$ is given by
\beq
P_i^j=\frac{2}{\sqrt{g}}\sin(\frac{\pi i j}{2g})\,\delta_{(i+j+g)\mod 2,0}
\label{eq:wzwP}
\eeq
and for $g$ even
\begin{eqnarray}
(-1)^{\frac{g+2}{2}}\sum_{\substack{{j=1}\\ j=4n+1}}^{g-1}
P_k^j =
\frac{1}{2}\frac{S_{g/2}^k}{S_1^k}\left\{ P_k^1+(-1)^{\frac{g+2}{2}} P_k^{g-1}\right\}
\label{eq:Pidentity1}  \\
(-1)^{\frac{g+2}{2}}\sum_{\substack{{j=1}\\ j=4n+3}}^{g-1}
P_k^j =
-\frac{1}{2}\frac{S_{g/2}^k}{S_1^k}\left\{ P_k^1-(-1)^{\frac{g+2}{2}} P_k^{g-1}\right\} 
\label{eq:Pidentity3}
\end{eqnarray}
with $S_{g/2}^k=(-1)^{\frac{k-1}{2}}\sqrt{\frac{2}{g}}$ for $k$ odd.
Note that both sides of \eqref{eq:Pidentity1} and \eqref{eq:Pidentity3} are zero
for $k$ even. 

\subsubsection{$A_{g-1}$}\label{sec:A_g}

In the $A_{g-1}$ case, there are two simple currents, namely, $1$ and $\sigma$, where
$\sigma=N_{g-1}=(\delta_{i,g-j})$.
\beq
n_{i1}{}^1=\delta_{i,1}\qquad\mbox{ and }\qquad
(\sigma\cdot n_{i})_{1}{}^1=n_{i\,\sigma(1)}{}^1=\delta_{i,g-1}
\eeq
so that
\beq
\Gamma_k^{(1)}=\frac{\eta}{\sqrt{S_1^k}} \, P_k^1
\qquad\mbox{ and }\qquad 
\Gamma_k^{(2)}=\frac{\eta}{\sqrt{S_1^k}} \, P_k^{g-1}
\eeq
where $\eta=\pm1$ is an overall sign and is undetermined. The non-twisted Klein bottle partition functions are
\beq
K^{(1)}_{(1,1)}(q)=\sum_{j=1}^{g-1}(-1)^{j-1}\,\chi_j(q)\qquad\mbox{ and }\qquad 
K^{(2)}_{(1,1)}(q)=\sum_{j=1}^{g-1}\chi_j(q)
\eeq

\subsubsection{$D_{2\ell+1}$}\label{sec:Dodd}

The parent graph of $D_{2\ell+1}$ is $A_{g-1}$ with Coxeter number $g=4\ell$. The $\Bbb Z_2$ automorphism
of $D_{2\ell+1}$ graph is endowed in the intertwiners 
$\sigma=n_{g-1}$.

For both $\kappa=1,2$ cases,
\beq
n_{i1}{}^1= n_{i\,\sigma(1)}{}^1=
\delta_{i,1}+\delta_{i,g-1}
\label{eq:Doddintertwiner}
\eeq
and 
\beq
\psi_1{}^k=\left\{
\begin{array}{cr}
(-1)^{\frac{k-1}{2}}\sqrt{2} S_1^k & \quad k\neq\frac{g}{2}\\
0 &k=\frac{g}{2}
\end{array}
\right.
\eeq
so that
\beq
\Gamma_k^{(\kappa)}=\frac{(-1)^{\frac{k-1}{2}}}{\sqrt{2 S_1^k}} \, \eta
( P_k^1+{\tilde\eta}^{(\kappa)} P_k^{g-1})\,,\qquad k\neq\frac{g}{2}
\label{eq:GammaDoddvacuum}
\eeq
and we need to determine the sign of ${\tilde\eta}^{(\kappa)}=\eta^{(\kappa)}_{g-1} \eta^{(\kappa)}_1$. 
Note that
$\Gamma_{\frac{g}{2}}^{(\kappa)}$ is not determined in \eqref{eq:GammaDoddvacuum}
since $\psi_1{}^{\frac{g}{2}}=0$ and \eqref{eq:Gammafromvacuum} does not apply.

Now we repeat the above procedure, not starting from the vacuum boundary state
but from the state $\frac{g}{2}^+$ (which corresponds to the node $2\ell+1$ in figure \ref{tbl:Graphs}).
The action of $\sigma$ on the nodes at the fork is 
$\sigma(\frac{g}{2}^{\pm})=\frac{g}{2}^{\mp}$, and
\beq
n_{i\frac{g}{2}^{+}}{}^{\!\frac{g}{2}^{+}}=\sum_{\substack{{j=1}\\{j=4n+1}}}^{g-1} \delta_{i,j}
\qquad\mbox{ and }\qquad
n_{i\frac{g}{2}^{-}}{}^{\!\frac{g}{2}^{+}}=\sum_{\substack{{j=1}\\{j=4n+3}}}^{g-1} \delta_{i,j}
\eeq
and
\beq
\psi_{\frac{g}{2}^+}{}^k=\left\{
\begin{array}{cr}
\frac{(-1)^{\frac{k-1}{2}}}{\sqrt{2}} S_{\frac{g}{2}}{}^k =\frac{1}{\sqrt{g}}& \quad k\neq\frac{g}{2}\\
\frac{1}{\sqrt{2}} &k=\frac{g}{2}
\end{array}
\right.
\eeq
Thus we have another expressions for $\Gamma_k^{(\kappa)}$
\beq
\Gamma_k^{(\kappa)}=\sqrt{g S_1^k} \sum_{\substack{j=1\\j=4n-1+2{\kappa}}}^{g-1}
\!\! P_k^{j}\,\eta_j^{(\kappa)}
\label{eq:GammaDoddnonvacuum}
\eeq
and it follows that $\Gamma_{\frac{g}{2}}^{(\kappa)}=0$ since $P_{\frac{g}{2}}^j=0$ for $j$ odd. 

Equating \eqref{eq:GammaDoddvacuum} and \eqref{eq:GammaDoddnonvacuum} and
comparing them with the identities \eqref{eq:Pidentity1} and  \eqref{eq:Pidentity3}, we arrive at
\begin{eqnarray}
\Gamma_k^{(1)}&=&\frac{(-1)^{\frac{k-1}{2}}}{\sqrt{2 S_1^k}} \, \eta\,
( P_k^1- P_k^{g-1})= -\eta \sqrt{g S_1^k} \sum_{\substack{j=1\\j=4n+1}}^{g-1} P_k^j \\
{}&=&(-1)^{\frac{(k+2)^2-1}{8}}\, \eta\,\frac{P_k^{\frac{g}{2}-1}}{\sqrt{S_1^k}}
\end{eqnarray}
and
\begin{eqnarray}
\Gamma_k^{(2)}&=&\frac{(-1)^{\frac{k-1}{2}}}{\sqrt{2 S_1^k}} \, \eta\,
( P_k^1 + P_k^{g-1})= \eta \sqrt{g S_1^k} \sum_{\substack{j=1\\j=4n+3}}^{g-1} P_k^j \\
{}&=&-(-1)^{\frac{(k+2)^2-1}{8}}\, \eta\,\frac{P_k^{\frac{g}{2}+1}}{\sqrt{S_1^k}}
\end{eqnarray}
The non-twisted partition functions are
\beq
K^{(1)}_{(1,1)}(q)=\sum_{\substack{j=1\\{j\text{ odd}}}}^{g-1}\chi_j(q)-\chi_{\frac{g}{2}}(q)\qquad\mbox{ and }\qquad 
K^{(2)}_{(1,1)}(q)=\sum_{\substack{j=1\\{j\text{ odd}}}}^{g-1}\chi_j(q)+\chi_{\frac{g}{2}}(q)
\eeq

\subsubsection{$D_{2\ell}$}\label{sec:Deven}

$D_{2\ell}$ is a Type I graph with Coxeter number $g=4\ell-2$.
As in the $D_{2\ell+1}$ case, $n_{i1}{}^1$ and $n_{i\sigma(1)}{}^1$ are given by \eqref{eq:Doddintertwiner}.
However, unlike the  $D_{2\ell+1}$ case, $\sigma$ is not endowed neither in the graph fusion matrices of
$D_{2\ell}$ nor in the intertwiners with $A_{g-1}$. In fact, $\sigma$ is the non-faithful nimrep of the $\Bbb Z_2$ simple current
of the Ocneanu algebra $\tilde D_{2\ell}$.

The coefficients $\psi_1^k$ are given by
\beq
\psi_1{}^k=\left\{
\begin{array}{cr}
\sqrt{2} S_1^k & \quad k\neq\frac{g}{2}\\
S_1^{\frac{g}{2}} &k=\frac{g}{2}
\end{array}
\right.
\eeq
so that
\begin{eqnarray}
\Gamma_k^{(\kappa)}&=&\frac{1}{\sqrt{2 S_1^k}} \, \eta
( P_k^1+{\tilde\eta}^{(\kappa)} P_k^{g-1})\,,\qquad k\neq{\frac{g}{2}}^{\pm} \\
\label{eq:GammaDevenvacuum}
\Gamma_{{\frac{g}{2}}^{+}}^{(\kappa)}+\Gamma_{{\frac{g}{2}}^{-}}^{(\kappa)}&=&\frac{1}{\sqrt{S_1^{\frac{g}{2}}}} \, \eta
( P_{\frac{g}{2}}^1+{\tilde\eta}^{(\kappa)} P_{\frac{g}{2}}^{g-1})
\end{eqnarray}
To determine the sign of ${\tilde\eta}^{(\kappa)}$ and the crosscap coefficients for $k={\frac{g}{2}}^{\pm}$,
we repeat the procedure from the boundary state $\frac{g}{2}^+$ as we did in the $D_{2\ell+1}$ case.
\beq
\psi_{\frac{g}{2}^+}{}^k=\left\{
\begin{array}{cr}
\frac{1}{\sqrt{2}} S_{\frac{g}{2}}{}^k =\frac{(-1)^{\frac{k-1}{2}}}{\sqrt{g}}& \quad k\neq\frac{g}{2}\\
\frac{1}{2}(\,S_{\frac{g}{2}}^{\,\frac{g}{2}}\pm i \sqrt{(-1)^{\frac{g+2}{4}}} \, )
 &k={\frac{g}{2}}^{\pm}
\end{array}
\right.
\eeq
and we obtain different expressions for the crosscap coefficients.
Comparing with the identities \eqref{eq:Pidentity1} and  \eqref{eq:Pidentity3}, we arrive at
the results
\begin{equation}
\Gamma_k^{(1)}=\frac{2^{-\frac{1}{2}\delta_{k,g/2}}}{\sqrt{2 S_1^k}} \, \eta\,
( P_k^1+P_k^{g-1})
= \eta (-1)^{\frac{k-1}{2}}2^{-\frac{1}{2}\delta_{k,g/2}}
\sqrt{g S_1^k} \sum_{\substack{j=1\\j=4n+1}}^{g-1} P_k^j 
\end{equation}
and
\begin{equation}
\Gamma_k^{(2)}=\frac{2^{-\frac{1}{2}\delta_{k,g/2}}}{\sqrt{2 S_1^k}} \, \eta\,
( P_k^1 - P_k^{g-1})
= -\eta (-1)^{\frac{k-1}{2}} 2^{-\frac{1}{2}\delta_{k,g/2}}
\sqrt{g S_1^k} \sum_{\substack{j=1\\j=4n+3}}^{g-1} P_k^j 
\end{equation}
The non-twisted partition functions are
\beq
K^{(1)}_{(1,1)}(q)=\sum_{\substack{j=1\\{j\text{ odd}}\\j\neq g/2}}^{g-1}\chi_j(q)+2\chi_{\frac{g}{2}}(q)
= \sum_{\substack{a=1\\{a\text{ odd}}\\ a\neq 2\ell-1}}^{2\ell}\hat\chi_a(q)+ 
\hat\chi_{2\ell-1}(q)+ \hat\chi_{2\ell}(q)
\eeq
\beq
K^{(2)}_{(1,1)}(q)=\sum_{\substack{j=1\\{j\text{ odd}}\\j\neq g/2}}^{g-1}\chi_j(q)
= \sum_{\substack{a=1\\{a\text{ odd}}\\ a\neq 2\ell-1}}^{2\ell}\hat\chi_a(q)+ 
\hat\chi_{2\ell-1}(q) - \hat\chi_{2\ell}(q)
\eeq
where $\hat \chi_a(q)$ are the extended characters defined as
\beq
\hat \chi_a(q)= \sum_{i\in A_{g-1}} n_{i1}{}^a \chi_i(q)
\eeq

Note that the $\kappa=2$ case is not discussed in \cite{PSS95a} since
the Klein bottle coefficients do not satisfy the fusion rules of the
extended chiral algebra, namely, 
\beq
\hat K_a \hat K_b \hat K_c >0 \qquad \mbox{if} \quad \hat N_{ab}{}^c\neq 0
\label{eq:Kleinbottlefusion}
\eeq
where $\hat K_a$ are the Klein bottle coefficients in terms of the
extended characters. The debate as to whether the constraint \eqref{eq:Kleinbottlefusion}
is necessary for the Klein bottle coefficients is not settled. Refer to \cite{SS03}
for more detail.

Finally, we remark that the $D_4$ model possesses $\Bbb S_3$  symmetry which generates
6 simple currents. However, it turns out that they do not satisfy integrality 
except for the cases we consider here.

\subsubsection{$E_6$}\label{sec:E6}

$E_6$ has two simple currents, the identity and $\sigma=\hat N_5$.
\beq
n_{i1}{}^1=\delta_{i,1}+\delta_{i,7}\qquad\mbox{ and }\qquad
n_{i\,\sigma(1)}{}^1=n_{i\,5}{}^1=\delta_{i,5}+\delta_{i,11}
\eeq
By integrality,
\beq
\Gamma^{(1)}_k=\frac{\sqrt{S_1^k}}{\psi_1^k} \,\eta\,(P_k^1-P_k^7)
\eeq
and
\beq
\Gamma^{(2)}_k=\frac{\sqrt{S_1^k}}{\psi_1^k} \,\eta\,(P_k^5+P_k^{11})
\eeq
The non-twisted Klein bottle partition functions are
\beq
K^{(1)}_{(1,1,1)}(q)=\chi_1-\chi_4+\chi_5+\chi_7-\chi_8+\chi_{11}
=\hat\chi_1+\hat\chi_5-\hat\chi_6
\eeq
and
\beq
K^{(2)}_{(1,1,1)}(q)=\chi_1+\chi_4+\chi_5+\chi_7+\chi_8+\chi_{11}
=\hat\chi_1+\hat\chi_5+\hat\chi_6
\eeq

\subsubsection{$E_7$}\label{sec:E7}

$E_7$ is a Type II graph whose parent graph is $D_{10}$. 
\beq
n_{i1}{}^1=\delta_{i,1}+\delta_{i,9}+\delta_{i,17}
\eeq
By integrality,
\beq
\Gamma^{(1)}_k=\frac{\sqrt{S_1^k}}{\psi_1^k} \,\eta\,(P_k^1+P_k^9+P_k^{17})
\eeq
and the non-twisted Klein bottle partition function is
\beq
K^{(1)}_{(1,1,1)}(q)=(\chi_1+\chi_{17})+(\chi_{5}+\chi_{13})
+(\chi_{7}+\chi_{11})+\chi_{9}
=\hat\chi_1+\hat\chi_5+\hat\chi_7+\hat\chi_9
\label{eq:KleinE7}
\eeq
where $\hat\chi$ in \eqref{eq:KleinE7} are the extended characters of the parent graph $D_{10}$.

\subsubsection{$E_8$}\label{sec:E8}

\beq
n_{i1}{}^1=\delta_{i,1}+\delta_{i,11}+\delta_{i,19}+\delta_{i,29}
\eeq
By integrality,
\beq
\Gamma^{(1)}_k=\frac{\sqrt{S_1^k}}{\psi_1^k} \,\eta\,
(P_k^1-P_k^{11}-P_k^{19}-P_k^{29})
\eeq
and the non-twisted Klein bottle partition function is
\beq
K^{(1)}_{(1,1,1)}(q)=(\chi_1+\chi_{11}+\chi_{19}+\chi_{29})
+(\chi_{7}+\chi_{13}+\chi_{17}+\chi_{23})
=\hat\chi_1+\hat\chi_7
\eeq

\subsection{Unitary minimal models}\label{sec:unitaryminimalmodels}

One can generalize the method in the previous section to minimal models. 
In particular, we are interested in the unitary cases whose
realizations in statistical mechanics are known. These will be discussed in
section \ref{sec:LatticeModels}.

The unitary minimal models are labeled by a pair of graphs $(A_{g-2},G)$.
The central charge is $c=1-\frac{6}{g(g-1)}$ and the conformal weights
are given by the Kac formula subject to the Kac symmetry
\beq
h_{r,s}\equiv h_{g-1-r,g-s}= \frac{[gr-(g-1)s]^2-1}{4g(g-1)}
\eeq
For later convenience, we define the set of Kac indices $\mathcal I$ to be
\beq
\mathcal I=\left\{ \begin{array}{ll}
\{(r,s)\ |\ r \mbox{ odd },\ 1\leq s\leq g-1 \}\,, &\qquad \mbox{$g$ even}\\
\{(r,s)\ | \ 1\leq r\leq g-2,\ s \mbox{ odd } \}\,,&\qquad \mbox{$g$ odd} 
\end{array} \right. 
\label{eq:Kacindices}
\eeq
and the set of exponents $\mathcal E$ 
\beq
\mathcal E =\left\{\begin{array}{ll}
\{(r,a)\ | \ r \mbox{ odd }, a\in \mbox{Exp}(G) \}\,, &\qquad \mbox{$g$ even}\\
\mathcal I\,,&\qquad \mbox{$g$ odd}
\end{array}\right.
\label{eq:unitaryExponents}
\eeq
Note that $g$ is always even for $D$ and $E$ cases.

\noindent$\bullet$ \textbf{Twisted boundary conditions on the torus}

On the torus, the twisted partition functions are labeled by the 
tensor product Ocneanu graph $A_{g-2}\otimes\tilde G$
\beq
(r,x)=(r,a,b,\kappa)\in(A_{g-2},H,\bar H,\Bbb Z_2)
\label{eq:unitarytwistedboundarylabel}
\eeq 
and they are given by
\begin{equation}
Z_{(r,a,b,\kappa)}(q)=\sum_{(r',s),(r'',\bar s)}
N_{r\,r'}^{(A_{g-2})}{}^{r''}
[P_{ab}^{(\kappa)}]_{s \bar s}\;
\chi_{(r',s)}(q)\,\chi_{(r'',\bar s)}(\overline{q}),\quad
\label{eq:unitaryPZtwistedPFs}
\end{equation}

\hfill

\noindent$\bullet$ \textbf{Cylinder boundary conditions}

The conformal boundary conditions for the unitary minimal models are labeled
by the nodes of  tensor product graph $(r,a)\in A_{g-1}\otimes G$
with the symmetry
\beq
(r,a)\sim(g-1-r,\gamma(a))\,,\quad\mbox{ where }
\gamma=\left\{
\begin{array}{cl}
\sigma_G &\mbox{for $G=A,D_{2\ell+1},E_6$}\\
1 & \mbox{otherwise} 
\end{array}\right.
\label{eq:unitaryboundarysymmetry}
\eeq
The boundary states are expressed as linear sums of the Ishibashi states
\beq
|(r,a)\!>= \sum_{(r',s')\in\mathcal E} \frac{\Psi_{(r,a)}^{\ (r',s')}}
{\sqrt{ S_{(r,s)}^{(1,1)}}}\, |r',s'\!\gg
\eeq 
where
%
%
\beq
\Psi_{(r,a)}^{\ (r',s')}=\left\{\begin{array}{ll}
\sqrt{2} S^{({g-1})}_r{}^{r'}\,\psi_a{}^{s'} &\qquad r,r'\mbox{ odd},\, s'\in \mbox{Exp}(G) \mbox{ for $g$ even}\\
S_{(r,s)}^{(r',s')} &\qquad (r,s),(r',s')\in\mathcal I \mbox{ for $g$ odd}
\end{array} \right.
\label{eq:unitaryPsi}
\eeq
and $S_{(r,s)}^{(r',s')}= (-1)^{(r+s)(r'+s')} \sqrt{2} S^{({g-1})}_r{}^{r'}\,S^{(g)}_s{}^{s'} $
is the modular matrix, where
$S^{({g-1})}$ and $S^{({g})}$ are the modular matrices of $SU(2)_k$ at levels $g-3$ and $g-2$ respectively.

The cylinder partition function with boundaries $(r_1,a)$ and $(r_2,b)$ is
\begin{eqnarray}
A_{(r_1,a)|(r_2,b)}(q) &=& \sum_{(r,s)\in\mathcal I} n_{rs;(r_1,a)}{}^{(r_2,b)}\, \chi_{(r,s)}(q)\\
{}&=& \sum_{r,s} N_{r\,r_1}^{(A_{g-2})}{}^{r_2}\, n_{s\,a}{}^b\, \chi_{(r,s)}(q)
\end{eqnarray}
in which the non-negative coefficients are given by
\begin{eqnarray}
 n_{rs;(r_1,a)}{}^{(r_2,b)}&=& 
 \sum_{(r',s')\in\mathcal E} S_{(r,s)}^{(r',s')}\, \frac{\Psi_{(r,a)}^{\ (r',s')}}{\sqrt{S_{(1,1)}^{(r',s')}}}
 \, \frac{\Psi_{(r,b)}^{\ (r',s')}}{\sqrt{S_{(1,1)}^{(r',s')}}} \\
  {}&=&N_{r\,r_1}^{(A_{g-2})}{}^{r_2}\, n_{s\,a}{}^b+ N_{g-1-r\ r_1}^{(A_{g-2})}{}^{r_2}\,
  n_{g-s\ a}{}^b
\end{eqnarray}
where $N_{r}^{(A_{g-2})}$ and $n_s$ are the fused adjacency matrices of $A_{g-2}$ and $G$ respectively.

\hfill

\noindent$\bullet$ \textbf{Klein bottle}

Under the symmetry \eqref{eq:unitaryboundarysymmetry},
on the tensor product graph $A_{g-2}\otimes G$,  there are two independent
$\Bbb Z_2$ automorphisms, namely, the identity and $\sigma=(1,\sigma_G)$.
These $\Bbb Z_2$ automorphisms corresponds to the simple currents of 
the tensor Ocneanu algebra  $A_{g-2}\otimes\tilde G$ subject
to quantum symmetries.

The Klein bottle partition functions are labeled by
$(r,a,b)\in (A_{g-2},H,H)$
\beq
K_{(r,a,b)}^{(\kappa)}(q) = \sum_{(r'',s'')\in\mathcal I} 
K_{(r'',s'');(r,a,b)}^{(\kappa)}\, \chi_{(r'',s'')}(q)
\eeq
where $K_{(r'',s'');(r,a,b)}^{(\kappa)}$ are integer coefficients
\beq
K_{(r'',s'');(r,a,b)}^{(\kappa)} = \sum_{(r',s')\in\mathcal E} 
S_{(r'',s'')}^{(r',s')}\, \Gamma_{(r,s)}^{(\kappa)}  \Gamma_{(r,s)}^{(\kappa)*}
\, \frac{S^{({g-1})}_r{}^{r'}}{S^{({g-1})}_1{}^{r'}}\, \frac{\psi_a^{(H) s'}}{\psi_1^{(H) s'}}
\left(\frac{\psi_b^{(H) s'}}{\psi_1^{(H) s'}}\right)^*
\eeq

\hfill

\noindent$\bullet$ \textbf{M\"obius strip}

%
%

For unitary minimal models with Coxeter number $g$,
the $|\mathcal I|\times |\mathcal I|$ modular matrix $P$ 
is the sum of the direct products of the modular matrices $P^{(g-1)}$ and $P^{(g)}$ of 
$SU(2)_{g-3}$ and $SU(2)_{g-2}$ 
\beq
P_{(r_1,s_1)}^{(r_2,s_2)}=
\sigma_{r1,s1}\,(\, \sigma_{r2,s2}\, P^{(g-1)}_{r_1}{}^{r_2}  P^{(g)}_{s_1}{}^{s_2}\, +\,
 \sigma_{g-1-r2,g-s2}\, P^{(g-1)}_{r_1}{}^{g-1-r_2} P^{(g)}_{s_1}{}^{g-s_2}\,)
 \label{eq:unitaryPmatrix}
\eeq
where
\beq
\sigma_{r,s}=\left\{ \begin{array}{cl}
-1\,,& \quad(r-s)\,\mod 4\equiv 2 \\
1\,,&\quad\mbox{otherwise}
\end{array}\right. 
\eeq
Note that for the choice of $\mathcal I$ as defined in \eqref{eq:Kacindices},
the first term of \eqref{eq:unitaryPmatrix} vanishes.

The M\"obius strip partition functions are labeled by $(r,a)\in (A_{g-2},G)$
\beq
M_{(r,a)}^{(\kappa)}(q) = \sum_{i\in\mathcal I} M_{(r'',s'')}^{(\kappa)}
{\tilde\chi}_{(r,s)}(q)
\eeq 
where $M_{(r'',s'')}^{(\kappa)}$ are the integer coefficients
\beq
M_{(r'',s'')}^{(\kappa)}=\sum_{(r',s')\in\mathcal E} P_{(r'',s'')}^{(r',s')}\,
\frac{\Psi_{(r,a)}^{\ (r',s')}}{\sqrt{S_{(1,1)}^{(r',s')}}}\,
\Gamma_{(r',s')}^{(\kappa)}
\eeq
In addition to the symmetry \eqref{eq:unitaryboundarysymmetry},
$(r,a)\sim (r,\gamma(a))$ since $\psi_{a}{}^j$ equals $\psi_{\gamma(a)}{}^j$
up to a sign \cite{BPPZ00}.

In the following we list the crosscap coefficients of the unitary \ade models 
and their non-twisted partition functions on the Klein bottle.

\subsubsection{$(A_{g-2},A)$} \label{sec:unitaryA}

\beq
\Gamma_{(r,s)}^{(1)}= \frac{\eta}{\sqrt{S_{(1,1)}^{(r,s)}}}\, P_{(r,s)}^{(1,1)}\,,\qquad
\Gamma_{(r,s)}^{(2)}= \frac{\eta}{\sqrt{S_{(1,1)}^{(r,s)}}}\, P_{(r,s)}^{(1,g-1)}
\eeq 
and
\beq
K^{(1)}_{(1,1,1)}(q)=\sum_{(r,s)\in\mathcal I} \chi_{(r,s)}(q)\,,\qquad
K^{(2)}_{(1,1,1)}(q)=\sum_{(r,s)\in\mathcal I} (-1)^{r+s}\chi_{(r,s)}(q)
\eeq

\subsubsection{$(A_{g-2},D_{2\ell+1})$} \label{sec:unitaryDodd}


\beq
\Gamma_{(r,s)}^{(\kappa)}=\frac{(-1)^{\frac{s-1}{2}}}{\sqrt{2 S_{(1,1)}^{(r,s)}}} \, \eta\,
( P_{(r,s)}^{(1,1)}-(-1)^{\kappa}\, P_{(r,s)}^{(1,g-1)})
\eeq
and
\beq
K^{(\kappa)}_{(1,1,1)}(q)=\sum_{\substack{r=1\\{r\text{ odd}}}}^{g-1}\,
\sum_{\substack{s=1\\{s\text{ odd}}}}^{g-1}\chi_{(r,s)}(q)-(-1)^{\kappa}\,
\chi_{r,\frac{g}{2}}(q)
\eeq

\subsubsection{$(A_{g-2},D_{2\ell})$} \label{sec:unitaryDeven}


\beq
\Gamma_{(r,s)}^{(\kappa)}=\frac{2^{-\frac{1}{2}\delta_{s,g/2}}}{\sqrt{2 S_{(1,1)}^{(r,s)}}} \, \eta\,
( P_{(r,s)}^{(1,1)}-(-1)^{\kappa}\, P_{(r,s)}^{(1,g-1)})
\eeq
and
\beq
K^{(1)}_{(1,1,1)}(q)=\sum_{\substack{r=1\\{r\text{ odd}}}}^{g-1}\,
\sum_{\substack{s=1\\{s\text{ odd}}\\s\neq g/2}}^{g-1}\chi_{(r,s)}(q)+2\chi_{r,\frac{g}{2}}(q)
\eeq
\beq
K^{(2)}_{(1,1,1)}(q)=\sum_{\substack{r=1\\{r\text{ odd}}}}^{g-1}\,
\sum_{\substack{s=1\\{s\text{ odd}}\\s\neq g/2}}^{g-1}\chi_{(r,s)}(q)
\eeq

\subsubsection{$(A_{10},E_6)$} \label{sec:unitaryE6}

\beq
\Gamma_{(r,s)}^{(\kappa)}=\frac{\sqrt{ S_{(1,1)}^{(r,s)}}} {\Psi_{(1,1)}^{\ (r,s)}}\, \eta\,
( P_{(r,s)}^{(1,1)}-(-1)^{\kappa}\, P_{(r,s)}^{(1,7)})
\eeq
and
\begin{eqnarray}
K^{(\kappa)}_{(1,1,1)}(q)&=&\sum_{\substack{r=1\\{r\text{ odd}}}}^{g-1}\,
(\chi_{(r,1)}+\chi_{(r,7)}) +(\chi_{(r,5)}+\chi_{(r,11)})-(-1)^{\kappa}(\chi_{(r,4)}+\chi_{(r,8)}) \\
{}&=&
\sum_{\substack{r=1\\{r\text{ odd}}}}^{g-1}\,
 {\hat\chi}_{(r,1)}(q)+{\hat\chi}_{(r,5)}(q)-(-1)^{\kappa} {\hat\chi}_{(r,6)}(q)
\end{eqnarray}

\subsubsection{$(A_{16},E_7)$}\label{sec:unitaryE7}

\beq
\Gamma_{(r,s)}^{(1)}=\frac{\sqrt{ S_{(1,1)}^{(r,s)}}} {\Psi_{(1,1)}^{\ (r,s)}}\, \eta\,
( P_{(r,s)}^{(1,1)}+ P_{(r,s)}^{(1,9)}+P_{(r,s)}^{(1,17)})
\eeq
and
\begin{eqnarray}
\!\!K^{(\kappa)}_{(1,1,1)}(q)&\!=\!&\sum_{\substack{r=1\\{r\text{ odd}}}}^{g-1}\,
(\chi_{(r,1)}+\chi_{(r,17)}) +(\chi_{(r,5)}+\chi_{(r,13)})+(\chi_{(r,7)}+\chi_{(r,11)})+\chi_{(r,9)} \\
\!\!{}&\!=\!&
\sum_{\substack{r=1\\{r\text{ odd}}}}^{g-1}\,
 {\hat\chi}_{(r,1)}(q)+{\hat\chi}_{(r,5)}(q)+{\hat\chi}_{(r,7)}(q)+{\hat\chi}_{(r,9)}(q)
\end{eqnarray}
where ${\hat\chi}_{(r,a)}(q)$ are the extended characters of the parent graph $(A_{16},D_{10})$.

\subsubsection{$(A_{28},E_8)$} \label{sec:unitaryE8}

\beq
\Gamma_{(r,s)}^{(1)}=\frac{\sqrt{ S_{(1,1)}^{(r,s)}}} {\Psi_{(1,1)}^{\ (r,s)}}\, \eta\,
( P_{(r,s)}^{(1,1)}+P_{(r,s)}^{(1,11)}+ P_{(r,s)}^{(1,19)}-P_{(r,s)}^{(1,29)})
\eeq
and
\begin{eqnarray}
\!\!\!\!\!\!\!\!K^{(\kappa)}_{(1,1,1)}(q)\!&\!=\!&\!\sum_{\substack{r=1\\{r\text{ odd}}}}^{g-1}\,
(\chi_{(r,1)}\!+\!\chi_{(r,11)}\! +\!\chi_{(r,19)}\!+\!\chi_{(r,29)})+(\chi_{(r,7)}\!+\!\chi_{(r,13)}\!+\!\chi_{(r,17)}\! +\!\chi_{(r,23)})\\
\!\!\!\!{}&\!=\!&
\sum_{\substack{r=1\\{r\text{ odd}}}}^{g-1}\,
 {\hat\chi}_{(r,1)}(q)+{\hat\chi}_{(r,7)}(q)
\end{eqnarray}

\section{Lattice Realizations of the Klein Bottle and M\"obius Strip Amplitudes}\label{sec:LatticeModels}

It is well known that the $s\ell(2)$ unitary minimal theories are realized as
the continuum scaling limit of critical  \ade lattice models \cite{Pas}. Theses solvable lattice models are associated with the Dynkin diagrams $G$ of 
a simply laced Lie algebra of $A$, $D$, or $E$ type. 
The spin states $a,b,c,d$ are nodes of the graph $G$ and the spins of adjacent sites on the lattice
must be adjacent nodes on the graph.
The (unfused) Boltzmann face weight is given by
\begin{equation}
    W^{11}\W{d&c\\a&b}{|u}=
    \setlength{\unitlength}{.4cm}
\begin{picture}(3.5,1)(-.6,-.2)
\put(0,-1){\framebox(2,2){\p{}{u}}}
\put(.1,-.9){\pp{bl}{\searrow}}
\put(-.25,-1.25){\pp{}{a}}
\put(2.25,-1.25){\pp{}{b}}
\put(2.25,1.25){\pp{}{c}}
\put(-.25,1.25){\pp{}{d}}
\end{picture}
=s(\lambda-u)\,\delta_{ac}+
    s(u)\,\sqrt{\frac{\psi_{a}\psi_{c}}{\psi_{b}\psi_{d}}}\;\delta_{bd}
           \label{eq:W11}
\end{equation}
 and zero otherwise.  Here, 
 $u$ is the spectral parameter with $0<u<\lambda$, 
 $\lambda=\frac{\pi}{g}$, $s(u)=\frac{\sin (u)}{\sin (\lambda)}$ and
 $\psi_a$ are the entries of the Perron-Frobenius eigenvector
of the adjacency matrix $G$.

The CFT on a torus with defect lines is realized as an insertion of seams on the lattice.
There are three kinds of seams, namely, $r$-type, $a$-type and automorphism seam.
Generally, the $r$-type and $a$-type seams are simply  fused Boltzmann weights.
The product of the seams are identically labelled by 
$(r,x)=(r,a,b,\kappa)\in(A_{g-2},H,\bar H,\Bbb Z_2)$ as in \eqref{eq:unitarytwistedboundarylabel},
where $r$ denotes the $r$-type seam, and $a$ and $b$ denote the respective left-chiral and right-chiral $a$-type seams, and $\kappa$ the automorphism seam. 
This composite seam realizes the corresponding twisted boundary conditions.

%

\subsection{Seam}\label{sec:Seam}

Generally, the $r$-type and $a$-type seams are simply  fused Boltzmann weights.
In this section, we quickly recall the definition of seams. For the details on the construction of seams,
we refer to \cite{BP01,CMP03}.

The $r$-type seam
${W_{(r,1)}}\smallW{d&\gamma&c\\ a&\alpha&b}{|\!u, \xi}$
is obtained by fusing $r-1$ faces
\setlength{\unitlength}{10mm}
\begin{equation}
\begin{picture}(8,3)
      \put(0,1.5){\pp{c}{{\ds W_{(r,1)}}\W{d&\gamma&c\\ 
a&\alpha&b}{|u,\xi}\;\;=\;\;
      {\ds \frac{1}{
      s^{r-2}_{-1}(u+\xi)}}
      }}
      \put(3.5,0.5){
\begin{picture}(5,2)
\multiput(0.5,0.5)(0,1){2}{\line(1,0){4}}
\multiput(0.5,0.5)(1,0){2}{\line(0,1){1}}
\multiput(3.5,0.5)(1,0){2}{\line(0,1){1}}
\multiput(0.48,0.58)(3,0){2}{\pp{bl}{\searrow}}
\put(4,1){\pp{c}{u+\xi}}
\put(2.5,1){\pp{c}{\cdots}}
\put(0.6,1.17){\pp{l}{u+\xi-}}\put(1.45,0.9){\pp{r}{(r-2)\lambda}}
\put(0.48,0.48){\pp{tr}{a}}\put(4.52,0.48){\pp{tl}{b}}
\put(0.48,1.52){\pp{br}{d}}\put(4.52,1.52){\pp{bl}{c}}
\put(1.5,0.39){\pp{t}{e_1}}\put(3.6,0.39){\pp{t}{e_{r-2}}}
\put(1.5,1.64){\pp{b}{g_1}}\put(3.6,1.64){\pp{b}{g_{r-2}}}
\multiput(1.5,0.5)(2,0){2}{\pp{}{\bullet}}
\multiput(1.5,1.5)(2,0){2}{\pp{}{\bullet}}
\end{picture}}
\put(6,0){\p{b}{U^r_{\alpha}(a,b)_{(a,e_1,\ldots,e_{r\mi2},b)}}}
\put(6,3){\p{t}{U^r_{\gamma}(d,c)^{*}_{(d,g_1,\ldots,g_{r\mi2},c)}}}
\end{picture}
      \label{eq:Wr1}
\end{equation}
where $s^p_i(u)=\prod_{j=0}^{p-1} s(u+(i-j)\lambda)$, and $\vec U^r(a,b)$
are the fusion vectors which are the normalized eigenvectors of the 
idempotent fusion projector $P^r(a,b)$ with non-zero eigenvalues
\cite{BP01}, and $\alpha= 1,2,\ldots,n_{r a}{}^b$ and
$\gamma=1,2,\ldots,n_{r c}{}^d$ are the bond variables labeling these
eigenvectors. 
The solid dots indicate that the spins are summed out.

The left-chiral $a$-type seam is the braid limit of the fused face weight
\setlength{\unitlength}{10mm}
\begin{equation}
\begin{picture}(8,3)
      \put(0,1.5){\pp{c}{{\ds W_{(1,a)}}\W{d&\gamma&c\\ 
p&\alpha&q}{.}\;\;=\;\;
      {\ds \lim_{\xi\rightarrow i\infty}}\,\ 
      {\ds \frac{\exp(-i
      \frac{(g+1)(s-1)}{2}\lambda
      )}{
     s(u+\xi)\,
      s^{r-2}_{-1}(u+\xi)}}
      }}
      \put(4,0.5){
\begin{picture}(5,2)
\multiput(0.5,0.5)(0,1){2}{\line(1,0){4}}
\multiput(0.5,0.5)(1,0){2}{\line(0,1){1}}
\multiput(3.5,0.5)(1,0){2}{\line(0,1){1}}
\multiput(0.48,0.58)(3,0){2}{\pp{bl}{\searrow}}
\put(4,1){\pp{c}{u+\xi}}
\put(2.5,1){\pp{c}{\cdots}}
\put(0.6,1.17){\pp{l}{u+\xi-}}\put(1.45,0.9){\pp{r}{(s-2)\lambda}}
\put(0.48,0.48){\pp{tr}{a}}\put(4.52,0.48){\pp{tl}{b}}
\put(0.48,1.52){\pp{br}{d}}\put(4.52,1.52){\pp{bl}{c}}
\put(1.5,0.39){\pp{t}{e_1}}\put(3.6,0.39){\pp{t}{e_{s-2}}}
\put(1.5,1.64){\pp{b}{g_1}}\put(3.6,1.64){\pp{b}{g_{s-2}}}
\multiput(1.5,0.5)(2,0){2}{\pp{}{\bullet}}
\multiput(1.5,1.5)(2,0){2}{\pp{}{\bullet}}
\end{picture}}
\put(6.5,0){\p{b}{\hat U^{(s)a}_{\alpha}(p,q)_{(p,e_1,\ldots,e_{s\mi2},q)}}}
\put(6.5,3){\p{t}{\hat U^{(s)a}_{\gamma}(d,c)^{*}_{(d,g_1,\ldots,g_{s\mi2},c)}}}
\end{picture}
      \label{eq:W1s}
\end{equation}
where $\hat U^{(s)a}(d,c)$ are the fusion vectors of the $\hat N$ fusion projector $\hat P^{(s)a}(d,c)$
which are obtained from decomposition of $P^s(d,c)$ \cite{CMP03}
\beq
P^s(d,c)=\sum_{a\in G} n_{s1}{}^a \hat P^{(s)a}(d,c)
\eeq
The right-chiral $a$-type seam is given by the braid limit $\xi\rightarrow -i\infty$.
Since the automorphism seam does not play a role on non-orientable topologies, we do not 
discuss it here.

\subsection{Torus and Klein bottle transfer matrices}\label{sec:torusxfer}

In this section, we construct transfer matrices on the Klein bottle.
We start by constructing a transfer matrix on a torus. A torus is obtained by
joining the ends of a cylinder. The boundaries are periodic. On the other hand,
the Klein bottle is achieved by twisting one end of the cylinder before gluing them
together. Correspondingly, for the partition function on the Klein bottle, 
a flip operator is inserted before taking the trace of the transfer matrix.
This transfer matrix must be consistent with the flip operator as we shall see below.

The entries of the single-row torus transfer matrix 
$\vec{T}_{(r,a)}(u,\xi)$
with a pair of $(r,a)$ seams inserted are
defined as
\begin{gather}
\rule{0pt}{24pt}
\< \vec{a},\vec{\alpha}|\;
\vec{T}_{(r,a)}(u,\xi)\;
       |\vec{b},\vec{\beta}\>=
      \hspace{7cm}
       \notag\\[16pt]
\setlength{\unitlength}{12mm}
\vtop to .7\unitlength {}
\begin{picture}(7.2,.8)(.4,.9)
\put(-2.5,0.8){\large $
 \sqrt{\frac{\psi_{a_1} \psi_{b_3}}{\psi_{a_3} \psi_{b_1}}}$ }
  \multiput(-0.8,0.5)(0,1){2}{\line(1,0){2}}
  \multiput(-0.8,0.5)(1,0){3}{\line(0,1){1}}
  \multiput(0,0)(1,0){2}{
\put(-0.8,0.52){\pp{bl}{\sss\searrow}}
}
\put(-0.25,1){\pp{}{{}^{r}\!(\lambda\!-\!u,\xi)}}
\put(0.75,1){\pp{}{{(1,a)}}}
\put(-0.8,0.35){\pp{c}{a_{3}}}
\put(-0.3,0.5){\pp{c}{\alpha_{2}}}
\put(0.2,0.35){\pp{c}{a_{2}}}
\put(0.7,0.5){\pp{c}{\alpha_{1}}}
\put(1.2,0.35){\pp{c}{a_{1}}}
\put(-0.8,1.65){\pp{c}{b_{3}}}
\put(-0.3,1.5){\pp{c}{\beta_{2}}}
\put(0.2,1.65){\pp{c}{b_{2}}}
\put(0.7,1.5){\pp{c}{\beta_{1}}}
\put(1.2,1.65){\pp{c}{b_{1}}}
\put(1.5,0.85){$\times$}
\put(1.8,0.9){
\begin{picture}(7.2,.8)(.4,.9)
     \multiput(0.5,0.5)(0,1){2}{\line(1,0){4}}
\multiput(0.5,0.5)(1,0){2}{\line(0,1){1}}
\multiput(3.5,0.5)(1,0){2}{\line(0,1){1}}
\put(0.5,0.52){\pp{bl}{\sss\searrow}}
\multiput(3,0)(1,0){1}{
\put(0.5,0.52){\pp{bl}{\sss\searrow}}
}
\put(2.5,1){\pp{}{\cdots}}
\put(0.5,0.35){\pp{c}{a_{3}}}
\put(1.5,0.35){\pp{c}{a_{4}}}
\put(3.5,0.35){\pp{c}{a_{N+2}}}
\put(4.5,0.35){\pp{c}{a_{N\plus3}}}
\put(0.5,1.7){\pp{c}{b_{1}}}
\put(1.5,1.7){\pp{c}{b_{2}}}
\put(3.5,1.7){\pp{c}{b_{N\plus2}}}
\put(4.5,1.7){\pp{c}{b_{N\plus3}}}
\put(1,1){\pp{}{u}}
\put(4,1){\pp{}{u}}
\put(2.1,0.8){\p{}{}}\end{picture}
}
\put(6.25,0.85){$\times$}
\put(7.9,0.87){
\begin{picture}(7.2,.8)(.4,.9)
  \multiput(-0.8,0.5)(0,1){2}{\line(1,0){2}}
  \multiput(-0.8,0.5)(1,0){3}{\line(0,1){1}}
  \multiput(0,0)(1,0){2}{
\put(-0.8,0.52){\pp{bl}{\sss\searrow}}
}
\put(-0.25,1){\pp{}{{}^{r}\!(u,\xi)}}
\put(0.75,1){\pp{}{{(1,a)}}}
\put(-0.8,0.35){\pp{c}{a_{N\plus3}}}
\put(-0.3,0.5){\pp{c}{\alpha_{N\plus3}}}
\put(0.2,0.35){\pp{c}{a_{N\plus4}}}
\put(0.7,0.5){\pp{c}{\alpha_{N\plus4}}}
\put(1.2,0.35){\pp{c}{a_{1}}}
\put(-0.8,1.65){\pp{c}{b_{N\plus3}}}
\put(-0.3,1.5){\pp{c}{\beta_{N\plus3}}}
\put(0.2,1.65){\pp{c}{b_{N\plus4}}}
\put(0.7,1.5){\pp{c}{\beta_{N\plus4}}}
\put(1.2,1.65){\pp{c}{b_{1}}}
\end{picture}
}
\end{picture}
       \notag \\[16pt]
=  \sqrt{\frac{\psi_{a_1} \psi_{b_3}}{\psi_{a_3} \psi_{b_1}}} \ 
    W_{(r,1)}\W{b_{3}&\beta_{2}&b_{2}\\
    {}\\
       a_{3}&\alpha_{2}&a_{2}}{|\lambda-u,\xi}\,
    W_{(1,a)}\W{b_{2}&\beta_{1}&b_{1}\\
    {}\\
    a_{2}&\alpha_{1}&a_{1}}{.}
    \times    \notag\\
   \prod_{i=3}^{N+2}W\W{b_{i}&{}&b_{i+1}\\
    {}\\
    a_{i}&{}&a_{i+1}}{|u}\times
    W_{(r,1)}\W{b_{N\plus 3}&\beta_{N\plus 3}&b_{N\plus 4}\\
    {}\\
    a_{N\plus 3}&\alpha_{N\plus 3}&a_{N\plus 4}}{|u,\xi}
    W_{(1,a)}\W{b_{N\plus 4}&\beta_{N\plus 4}&b_{1}\\
   {}\\
    a_{N\plus 4}&\alpha_{N\plus 4}&a_{1}}{.}
            \label{eq:singlerowTorusXfer}
\end{gather}
Note that the path of the left-hand seams is reversed. 
It will be clear
shortly that such construction is necessary in order to make the transfer
matrix consistent with the Klein bottle geometry.
One can show by crossing symmetry that the flipped $a$-type seam on the left hand side
is equivalent to the complex conjugate of a non-flipped $a$-type seam. This is the reason
the corresponding partition function is ambichiral.  
The transfer matrix forms a commuting family thanks to the generalized Yang-Baxter equation and thus the lattice model is integrable.

The flip operator $\vec F$ which interchanges spins on the left and on the right
is given by
\beq
\< \vec{a},\vec{\alpha}|\;
\vec{F}\;
       |\vec{b},\vec{\beta}\>=
        \delta_{\alpha_{1},\beta_{N+4}}\,  \delta_{\alpha_{2},\beta_{N+3}}\,
        \delta_{\alpha_{N+3},\beta_{2}}\,  \delta_{\alpha_{N+4},\beta_{1}}\
       \prod_{i=1}^{N+4} \delta_{a_i,b_{N+6-i}}
\eeq
and the $\Bbb Z_2$ symmetry operator $\vec R$
\beq
\< \vec{a},\vec{\alpha}|\;
\vec R\;
       |\vec{b},\vec{\beta}\>=
        \delta_{\alpha_{1},\beta_{1}}\,  \delta_{\alpha_{2},\beta_{2}}\,
        \delta_{\alpha_{N+3},\beta_{N+3}}\,  \delta_{\alpha_{N+4},\beta_{N+4}}\
       \prod_{i=1}^{N+4} \delta_{a_i,\sigma(b_{i})}
\eeq
i.\,e.\ the $\Bbb Z_2$ automorphism acts on spins at each site locally.  
Clearly $\vec F$ and $\vec R$ commute and $\vec F^2=\vec R^2=1$.

One can show that
\beq
\vec F\, \vec T_{(r,a)}(u,\xi) =  \vec T_{(r,a)}(\lambda-u,\xi)\, \vec F
\label{eq:FTcommute}
\eeq
Note that in order for \eqref{eq:FTcommute} to hold, the flip of the left-hand seam and the existence of the
gauge factor in \eqref{eq:singlerowTorusXfer} are necessary.
Moreover, $[\vec R,\vec T_{(r,a)}]=0$ with an appropriate normalization of the fusion vectors
\beq
U^{(s)a}_{\alpha}(d,c)_{(d,g_1,\ldots,g_{s\mi2},c)}\,=\,
\hat U^{(s)a}_{\alpha}(\sigma(d),\sigma(c))_{(\sigma(d),\sigma(g_1),\ldots,\sigma(g_{s\mi2}),\sigma(c))}
\eeq
Such normalization is always possible since the adjacency condition
$G_{a,b}=G_{\sigma(a),\sigma(b)}$ and 
the Boltzmann weight is invariant under
the action of the $\Bbb Z_2$ automorphism on the spins
\beq
   W^{11}\W{d&c\\a&b}{|u}\,=\,
      W^{11}\W{\sigma(d)&\sigma(c)\\\sigma(a)&\sigma(b)}{|u}
\eeq

The double-row transfer matrix is defined as
\beq
\vec D_{(r,a)}(u,\xi) = \vec T_{(r,a)}(u,\xi)\, \vec T_{(r,a)}(\lambda-u,\xi)
\eeq
so that $\vec  F$ and $\vec D_{(r,a)}(u,\xi)$ commute for any value
of $u$ and $\xi$. 

We claim that the partition functions on the Klein bottle are given by
\footnote{\eqref{eq:Kleinpartitionfunction} is not valid for $A_{\mbox{\scriptsize even}}$.
See \eqref{eq:KleinpartitionAeven} in section \ref{sec:Computation}.}
\beq
K^{(\kappa)}_{(r,a)\,MN}(u,\xi)=\mbox{Tr}\, \vec F^{(\kappa)}\, (\vec D_{(r,a)}(u,\xi))^M\,,
\qquad
a\neq2\ell\!-\!1,2\ell \mbox{ for } D_{2\ell} \mbox{ when } \kappa=2
\label{eq:Kleinpartitionfunction}
\eeq
where $\vec F^{(\kappa)}=\vec F\vec R^{\kappa-1}$ and
$M$ is odd.
For $\kappa=1$,  $\vec F^{(1)}$ is  simply a geometric flip in the usual sense. On the other hand,
for $\kappa=2$, it involves a flip in the spin configuration space. This flip corresponds
to the $\Bbb Z_2$ symmetry of the Dynkin diagrams $G$. Such spin-state flip for the Ising 
model on non-orientable surfaces was known and 
its interpretation was given in \cite{FRS04}. 
The correctness of \eqref{eq:Kleinpartitionfunction} is verified numerically in section \ref{sec:Computation}.


\subsubsection{$D_{2\ell}$ case}\label{sec:KleinD2l}

The continuum scaling limit of the transfer matrix $\vec T_{(r,a)}$, with a pair of $(1,a)$ seams inserted  
as in \eqref{eq:singlerowTorusXfer}, gives the conformal partition function
$Z_{(1,a,a,1)}(q)$ on a torus.
For the $D_{2\ell}$ cases, whose Ocneanu algebras are non-commutative,  
the quantum symmetries of $D_{2\ell}$ for the nodes at the fork of the graph $G$ give
\beq
Z_{(r,2\ell-1,\kappa)}=
\left\{
\begin{array}{ll}
Z_{(r,1,2\ell-1,\kappa)} & \qquad\ell \mbox{ odd} \\
Z_{(r,1,2\ell,\kappa)} & \qquad\ell \mbox{ even}
\end{array}
\right.
\eeq 
and similarly for the node $2\ell$.
This commutation relation is reflected on the graph fusion matrices
\beq
\begin{array}{ll}
\hat N_{2\ell} \mbox{ and } \hat N_{2\ell-1} \mbox{ symmetric } &\qquad
\ell \mbox{ odd} \\
\hat N_{2\ell}=\hat N_{2\ell-1}^T & \qquad\ell \mbox{ even}
\end{array}
\label{eq:D2lNhatrelation}
\eeq
and $\sigma\,\hat N_{2\ell-1}=  \hat N_{2\ell-1}\, \sigma$.

On the Klein bottle, the partition function \eqref{eq:Kleinpartitionfunction} is not defined for
$D_{2\ell}$ when $\ell$ is even, since the spin path is not admissible
under the action of $\vec F^{(2)}$.  

To realize the partition function for $K^{(2)}_{(r,2\ell)}(q)=K^{(2)}_{(r,2\ell-1)}(q)$,
the partition function for the lattice model is given by
\beq
K^{(2)}_{(r,2\ell)\,MN}(u,\xi) = \mbox{Tr}\,(\vec F^{(2)}\, 
(\vec T'_{r,2\ell}(u,\xi)\, \vec T'_{r,2\ell}(\lambda-u,\xi))^M)
\eeq
where $\vec T'_{r,2\ell}(u,\xi)$ is defined as in \eqref{eq:singlerowTorusXfer}
but with $(1,2\ell)$ and $(1,2\ell-1)$ seams at the left and right ends instead
\begin{gather}
\rule{0pt}{24pt}
\< \vec{a},\vec{\alpha}|\;
\vec T'_{(r,2\!\ell)}(u,\xi)\;
       |\vec{b},\vec{\beta}\>=
      \hspace{7cm}
       \notag\\[16pt]
\setlength{\unitlength}{12mm}
\vtop to .7\unitlength {}
\begin{picture}(7.2,.8)(.4,.9)
\put(-2.5,0.8){\large $
 \sqrt{\frac{\psi_{a_1} \psi_{b_3}}{\psi_{a_3} \psi_{b_1}}}$ }
  \multiput(-0.8,0.5)(0,1){2}{\line(1,0){2}}
  \multiput(-0.8,0.5)(1,0){3}{\line(0,1){1}}
  \multiput(0,0)(1,0){2}{
\put(-0.8,0.52){\pp{bl}{\sss\searrow}}
}
\put(-0.25,1){\pp{}{{}^{r}\!(\lambda\!-\!u,\xi)}}
\put(0.75,1){\pp{}{{(1,2\ell)}}}
\put(-0.8,0.35){\pp{c}{a_{3}}}
\put(-0.3,0.5){\pp{c}{\alpha_{2}}}
\put(0.2,0.35){\pp{c}{a_{2}}}
\put(0.7,0.5){\pp{c}{\alpha_{1}}}
\put(1.2,0.35){\pp{c}{a_{1}}}
\put(-0.8,1.65){\pp{c}{b_{3}}}
\put(-0.3,1.5){\pp{c}{\beta_{2}}}
\put(0.2,1.65){\pp{c}{b_{2}}}
\put(0.7,1.5){\pp{c}{\beta_{1}}}
\put(1.2,1.65){\pp{c}{b_{1}}}
\put(1.5,0.85){$\times$}
\put(1.8,0.9){
\begin{picture}(7.2,.8)(.4,.9)
     \multiput(0.5,0.5)(0,1){2}{\line(1,0){4}}
\multiput(0.5,0.5)(1,0){2}{\line(0,1){1}}
\multiput(3.5,0.5)(1,0){2}{\line(0,1){1}}
\put(0.5,0.52){\pp{bl}{\sss\searrow}}
\multiput(3,0)(1,0){1}{
\put(0.5,0.52){\pp{bl}{\sss\searrow}}
}
\put(2.5,1){\pp{}{\cdots}}
\put(0.5,0.35){\pp{c}{a_{3}}}
\put(1.5,0.35){\pp{c}{a_{4}}}
\put(3.5,0.35){\pp{c}{a_{N+2}}}
\put(4.5,0.35){\pp{c}{a_{N\plus3}}}
\put(0.5,1.7){\pp{c}{b_{1}}}
\put(1.5,1.7){\pp{c}{b_{2}}}
\put(3.5,1.7){\pp{c}{b_{N\plus2}}}
\put(4.5,1.7){\pp{c}{b_{N\plus3}}}
\put(1,1){\pp{}{u}}
\put(4,1){\pp{}{u}}
\put(2.1,0.8){\p{}{}}\end{picture}
}
\put(6.25,0.85){$\times$}
\put(7.9,0.87){
\begin{picture}(7.2,.8)(.4,.9)
  \multiput(-0.8,0.5)(0,1){2}{\line(1,0){2}}
  \multiput(-0.8,0.5)(1,0){3}{\line(0,1){1}}
  \multiput(0,0)(1,0){2}{
\put(-0.8,0.52){\pp{bl}{\sss\searrow}}
}
\put(-0.25,1){\pp{}{{}^{r}\!(u,\xi)}}
\put(0.75,1){\pp{}{{(1,2\ell\!-\!1)}}}
\put(-0.8,0.35){\pp{c}{a_{N\plus3}}}
\put(-0.3,0.5){\pp{c}{\alpha_{N\plus3}}}
\put(0.2,0.35){\pp{c}{a_{N\plus4}}}
\put(0.7,0.5){\pp{c}{\alpha_{N\plus4}}}
\put(1.2,0.35){\pp{c}{a_{1}}}
\put(-0.8,1.65){\pp{c}{b_{N\plus3}}}
\put(-0.3,1.5){\pp{c}{\beta_{N\plus3}}}
\put(0.2,1.65){\pp{c}{b_{N\plus4}}}
\put(0.7,1.5){\pp{c}{\beta_{N\plus4}}}
\put(1.2,1.65){\pp{c}{b_{1}}}
\end{picture}
}
\end{picture}
       \notag \\[16pt]
=  \sqrt{\frac{\psi_{a_1} \psi_{b_3}}{\psi_{a_3} \psi_{b_1}}} \ 
    W_{(r,1)}\W{b_{3}&\beta_{2}&b_{2}\\
    {}\\
       a_{3}&\alpha_{2}&a_{2}}{|\lambda-u,\xi}\,
    W_{(1,2\ell)}\W{b_{2}&\beta_{1}&b_{1}\\
    {}\\
    a_{2}&\alpha_{1}&a_{1}}{.}
    \times    \notag\\
   \prod_{i=3}^{N+2}W\W{b_{i}&{}&b_{i+1}\\
    {}\\
    a_{i}&{}&a_{i+1}}{|u}\times
    W_{(r,1)}\W{b_{N\plus 3}&\beta_{N\plus 3}&b_{N\plus 4}\\
    {}\\
    a_{N\plus 3}&\alpha_{N\plus 3}&a_{N\plus 4}}{|u,\xi}
    W_{(1,2\ell\!-\!1)}\W{b_{N\plus 4}&\beta_{N\plus 4}&b_{1}\\
   {}\\
    a_{N\plus 4}&\alpha_{N\plus 4}&a_{1}}{.}
            \label{eq:singlerowTorusXferDeven}
\end{gather}
so that it is consistent with the flip $F^{(2)}$.

\subsection{Cylinder and M\"obius strip transfer matrices}\label{sec:CylinMobXfer}

The vacuum boundary condition corresponds to $(r,a)=(1,1)$. 
The $(1,a)$ boundary weights, for two adjacent nodes of
$G$, $c$ and $a$ (i.\,e. $G_{ac}\neq 0$) are given by
\begin{gather}
    B_{(1,a)}\B{&&a\\
    {}&{}\\
       c\\
       {}&{}\\
                &&a}{.}=
\setlength{\unitlength}{8mm}
\begin{picture}(1.4,1)(-.25,.9)
\put(0,1){\line(1,-1){1}}
\put(0,1){\line(1,1){1}}
\multiput(1,0)(0,.205){10}{\line(0,1){.15}}
\put(.6,1){\pp{}{(1,a)}}
\put(1.1,2){\pp{l}{a}}
\put(0,1){\pp{r}{c\,}}
\put(1.1,0){\pp{l}{a}}
\end{picture}
= {\frac{\psi_{c}^{\half}}{\psi_{a}^{\half}}} \;
                \label{eq:B1a}
\end{gather}

\smallskip\noindent

An $(r,a)$ boundary weight is 
given by the action of $r$-type seams on the $(1,a)$-boundary
weight
\begin{gather}
    B_{(r,a)}\B{&&d&\; \delta\\
    {}&{}\\
       c\\
    {}&{}\\
                &&b&\; \beta}{|u,\xi}=
\setlength{\unitlength}{9mm}
\begin{picture}(1.8,1.2)(-.25,.9)
\put(1,2){\line(1,0){.5}}
\put(0,1){\line(1,-1){1}}
\put(0,1){\line(1,1){1}}
\put(1,0){\line(1,0){.5}}
\multiput(1.5,0)(0,.205){10}{\line(0,1){.15}}
\put(1,1.2){\pp{}{(r,a)}}
\put(1,.8){\pp{}{(u,\xi)}}
\put(1.5,2){\pp{l}{\,a}}
\put(1.25,2){\pp{}{\delta}}
\put(1,2.1){\pp{b}{d}}
\put(0,1){\pp{r}{c\;}}
\put(1,-.1){\pp{t}{b}}
\put(1.25,0){\pp{}{\beta}}
\put(1.5,0){\pp{l}{\,a}}
\end{picture}
\;= 
\;
\setlength{\unitlength}{9mm}
\begin{picture}(3.4,1)(-2.8,.9)
\put(-2.5,0){
\put(0,0){\framebox(2.5,2){}} 
\put(0,1){\line(1,0){2.5}}
\multiput(0,0)(0,1){2}{
\put(0.,.05){\pp{bl}{\sss\searrow}}
}}
\put(0,1){\line(1,-1){1}}
\put(0,1){\line(1,1){1}}
\multiput(1,0)(0,.205){10}{\line(0,1){.15}}
\multiput(.2,2)(.2,0){4}{\pp{}{.}}
\multiput(.2,0)(.2,0){4}{\pp{}{.}}
\put(-1.25,1.6){\pp{}{{}^{r}\!(\lambda\mi u,\, \xi)}}
\put(-1.25,.5){\pp{}{{}^{r}\!(u,\xi)}}
\put(.6,1){\pp{}{(1,a)\,}}
\put(1.1,2){\pp{l}{a}}
\put(-1.25,2){\pp{r}{\delta}}
\put(-2.5,2.1){\pp{b}{d}}
\put(-2.5,1){\pp{r}{c\;\;{}}}
\put(-2.5,-.1){\pp{t}{b}}
\put(-1.25,0){\pp{r}{\beta}}
\put(1.1,0){\pp{l}{a}}
\end{picture}
        \label{eq:Bra}
\end{gather}

\medskip\noindent
and the left boundary weights are equal to the right boundary up to a gauge factor 
\beq
 B_{(r,a)}\B{\delta\;&d&&\\
       {}&{}&{}\\ 
       &&& c\\
       {}&{}&{}\\ 
        \beta\;&b&& }{|u,\xi}=\,
\sqrt{\frac{\psi_d}{\psi_b}}
\,
 B_{(r,a)}\B{&&d&\; \delta\\
        {}&{}\\
       c\\
        {}&{}\\
                &&b&\; \beta}{|u,\xi}               
\label{eq:leftBra}
\eeq
These boundary weights satisfy the boundary Yang-Baxter equation.

The double row transfer matrix is given by
\begin{gather}
        \< \vec{a},\vec{\alpha}|\;
\vec{T}_{
        (r_{L},a_{L})|
                (r_{R},a_{R})
        }
        (u,\xi_{L},\xi_{R})\;
                |\vec{b},\vec{\beta}\>=\notag\\
\setlength{\unitlength}{14mm}
\vtop to 1.7\unitlength {}
\begin{picture}(9.5,1.7)(-.75,1.5)
     \multiput(0.5,0.5)(0,1){3}{\line(1,0){6.5}}
\multiput(0.5,0.5)(1,0){2}{\line(0,1){2}}
\multiput(5,0.5)(1,0){3}{\line(0,1){2}}
\multiput(0,0)(0,1){2}{
\put(0.5,.52){\pp{bl}{\sss\searrow}}}
\multiput(4.5,0)(1,0){2}{\put(0.5,1.52){\pp{bl}{\sss\searrow}}}
\multiput(4.5,0)(1,0){2}{\put(0.5,0.52){\pp{bl}{\sss\searrow}}}
\put(-1,0.35){\pp{c}{a_{L}}}
\put(-.75,0.5){\pp{c}{\alpha_{L}}}
\put(0.5,0.35){\pp{c}{a_{1}}}
\put(1.5,0.35){\pp{c}{a_{2}}}
\put(6,0.35){\pp{c}{a_{N}}}
\put(7,0.35){\pp{c}{a_{N\plus1}}}
\put(8.25,0.5){\pp{c}{\alpha_{R}}}
\put(8.5,0.35){\pp{c}{a_{R}}}
\put(-1,2.7){\pp{c}{a_{L}}}
\put(-.75,2.5){\pp{c}{\beta_{L}}}
\put(0.5,2.7){\pp{c}{b_{1}}}
\put(1.5,2.7){\pp{c}{b_{2}}}
\put(6,2.7){\pp{c}{b_{N}}}
\put(7,2.7){\pp{c}{b_{N\plus1}}}
\put(8.25,2.5){\pp{c}{\beta_{R}}}
\put(8.5,2.7){\pp{c}{a_R}}
\multiput(3,1)(0.25,0){4}{\pp{}{.}}
\multiput(3,2)(0.25,0){4}{\pp{}{.}}
%
\put(1,1){\p{}{u}}
\put(5.5,1){\p{}{u}}
\put(6.5,1){\p{}{u}}
\put(1,2){\p{}{\lambda\mi u}}
\put(5.5,2){\p{}{\lambda\mi u}}
\put(6.5,2){\p{}{\lambda\mi u}}
\put(-1,2.5){\line(1,0){.5}}
\put(-.5,2.5){\line(1,-1){1}}
\put(-.5,.5){\line(1,1){1}}
\put(-1,.5){\line(1,0){.5}}
\multiput(-1,.5)(0,.205){10}{\line(0,1){.15}}
\put(-.5,1.7){\pp{}{(r_{L},a_{L})}}
\put(-.5,1.3){\pp{}{(\lambda\mi u,\xi_{L})}}
\multiput(0,0)(0,2){2}{
        \multiput(0,0)(7.5,0){2}{
\multiput(.4,.5)(-.1,0){9}{\pp{}{.}}
        }}
\put(8.5,2.5){\line(-1,0){.5}}
\put(7,1.5){\line(1,-1){1}}
\put(7,1.5){\line(1,1){1}}
\put(8.5,.5){\line(-1,0){.5}}
\multiput(8.5,.5)(0,.205){10}{\line(0,1){.15}}
\put(8,1.7){\pp{}{(r_{R},a_{R})}}
\put(8,1.3){\pp{}{(u,\xi_{R})}}
\end{picture}
        \label{eq:TpqCylinder}
\end{gather}
One can show that $\vec T$ is symmetric.

Similarly, the flip operator $\vec F$ on the cylinder  
is given by
\beq
\< \vec{a},\vec{\alpha}|\;
\vec{F}\;
       |\vec{b},\vec{\beta}\>=
       \delta_{\alpha_{L},\beta_{R}}\,  \delta_{\alpha_{R},\beta_{L}}\
       \prod_{i=1}^{N+1} \delta_{a_i,b_{N+2-i}}
\eeq
and the $\Bbb Z_2$ symmetry operator $\vec R$
\beq
\< \vec{a},\vec{\alpha}|\;
\vec R\;
       |\vec{b},\vec{\beta}\>=
        \delta_{\alpha_{L},\beta_{L}}\,  \delta_{\alpha_{R},\beta_{R}}\
        \prod_{i=1}^{N+1} \delta_{a_i,\sigma(b_{i})}
\eeq
and $\vec F^{(\kappa)}=\vec F \vec R^{\kappa-1}$.

The M\"obius strip partition functions are
\beq
M^{(\kappa)}_{(r,a)\,MN}= \mbox{Tr}\, \vec F^{(\kappa)}(\vec T_{(r,\sigma^{\kappa-1}(a))|(r,a)}(u,\xi,\xi) )^M
\label{eq:Moebiuspartition}
\eeq
where $M$ is odd.

\section{Numerical Results}\label{sec:Computation}

The numerics described in this section were carried out using \textit{Mathematica}~\cite{Wolfram}.

The conformal points are  $(u,\xi)=(\frac{\lambda}{2},\xi_c)$, where
$\xi_c=\frac{\lambda}{2}(r-2+kg)$ for $k$ odd. 
At these points, $\vec F^{(\kappa)}$ and $\vec T_{(r,a)}$ commute.
Instead of diagonalizing the double-row transfer matrix, we diagonalize
$\vec F^{(\kappa)}\vec T_{(r,a)}(\frac{\lambda}{2},\xi_c)$. 
The reason for doing this is that
taking a product of large matrices in  \textit{Mathematica} consumes substantial memory.
For the same reason, we do not multiply the matrices $\vec F^{(\kappa)}$ and $\vec T_{(r,a)}$
directly.
Notice that the action of $\vec F^{(\kappa)}$ on $\vec T_{(r,a)}$ is simply
a permutation of the path basis of the transfer matrix. More precisely,
%
%
\beq
\< \vec{a},\vec{\alpha}|\;
\vec F^{(\kappa)}\vec T_{(r,a)}
\;
       |\vec{b},\vec{\beta}\>=
\< F^{(\kappa)}(\vec{a},\vec{\alpha})|\;
\vec T_{(r,a)}
\;
       |\vec{b},\vec{\beta}\>
\eeq
where $F^{(\kappa)}(\vec{a},\vec\alpha)$ denotes the permutation on the path 
\beq
\begin{array}{c}
(\vec a,\vec \alpha)=(a_1,a_2,a_3,\cdots,a_{N+3},a_{N+4},a_{1};\,\alpha_1,\alpha_2,\alpha_3,\alpha_4) \\
F^{(\kappa)}(\vec{a},\vec\alpha)=
(\sigma^{\kappa-1}(a_1),\sigma^{\kappa-1}(a_{N+4}),\sigma^{\kappa-1}(a_{N+3}),\cdots,\qquad\mbox{ }\\
\qquad\qquad\qquad\quad
\sigma^{\kappa-1}(a_{3}),\sigma^{\kappa-1}(a_{2}),\sigma^{\kappa-1}(a_{1});\,\alpha_4,\alpha_3,\alpha_2,\alpha_1)
\end{array}
\eeq

We project $ \vec T_{(r,a)}(\frac{\lambda}{2},\xi_c)$ onto the eigenbasis of $\vec F^{(\kappa)}$:
\beq
F^{(\kappa)}=\pm1\,:\qquad\qquad
\frac{1}{2}(\vec F^{(\kappa)}\pm1)\,  \vec T_{(r,a)}(\frac{\lambda}{2},\xi_c)
\label{eq:projectXfer}
\eeq
and diagonalize them separately. Thus, the partition function becomes
\beq
K^{(\kappa)}_{(r,a)\,MN}=\left(\sum_i |T^+_i|^{2M}\right) -
\left(\sum_j |T^-_j|^{2M}\right) 
\label{eq:eigensumXfer}
\eeq
where $T^\pm_i$ are the eigenvalues of \eqref{eq:projectXfer}. Since only the 
diagonal terms of the partition function survive on the Klein bottle, all
the non-diagonal terms cancel in \eqref{eq:eigensumXfer}. As we showed in \cite{CP02},
the non-diagonal terms have the characteristic that the zeros of the corresponding
eigenvalue functions are not symmetric on the upper and lower half complex plane.
Note that for the $A_{\mbox{\scriptsize even}}$ models, the trace of $\vec F^{(2)}\vec D_{(r,a)}$
is zero, since the action of $\Bbb Z_2$ automorphism $\sigma$ interchanges the sublattices
of even and odd spins. In these cases, the partition functions are defined as 
\beq
K^{(2)}_{(r,a)\,MN}=\sum_i (T^+_i)^{M} + \sum_j (T^-_j)^{M}\,,\qquad M\mbox{ odd}
\label{eq:KleinpartitionAeven}
\eeq

For $0<u<\lambda$, a critical lattice model gives rise to a conformal field theory in the continuum scaling
limit, namely, an $sl(2)$ unitary minimal model. This is manifest in the finite-size corrections to 
the eigenvalues of the transfer matrices.

For the periodic row transfer
matrix $\vec T_{(r,a)}(u,\xi)$ with a pair of seam $(r,a)$  and $N$ faces excluding the seams
as defined in \eqref{eq:singlerowTorusXfer},
the eigenvalues are 
\begin{equation}
T_n(u)=\exp(-E_n(u)),\quad n=0,1,2,\ldots
\end{equation}
The finite-size corrections to the energies $E_n$ take the form
\begin{eqnarray}
&&\hspace{1.7in}\mbox{} E_n(u)=Nf(u)+f_{r}(u,\xi) \label{eq:finiteSizeCorr}\\
&&\mbox{}\hspace{-.3in}\mbox{}
+\frac{2\pi}{N}\,\left(\big(-{\frac c{12}}+h_n+\bar h_n+k_n+\bar k_n\big)\sin\vartheta
+i(h_n-\bar h_n+k_n-\bar k_n)\cos\vartheta\right)
+o\left(\frac{1}{N}\right)\quad\mbox{}\notag
\end{eqnarray}
where $f(u)$ is the bulk free energy, $f_{r}(u,\xi)$ is the boundary
free energy due to the $r$-type seam, $c$ is the central
charge, $h_n$ and $\bar h_n$ are conformal weights,
$ k_n,\bar k_n\in{\Bbb N}$ label descendent levels and
the anisotropy angle $\vartheta$ is given by
\begin{equation}
\vartheta=g u
\end{equation}
where $g$ is the Coxeter number.

In the continuum scaling limit $M\gg N\gg 1$,
the conformal partition $K^{(\kappa)}_{(r,a)}(q)$ is given by
\beq
 K^{(\kappa)}_{(r,a)}(q) \simeq
\exp(Nf(u)+f_r(u,\xi_c))\,
 K^{(\kappa)}_{(r,a)\,MN}(u,\xi_c)
\eeq
where 
$q=\exp(4\pi i \tau)
\equiv q_{\mbox{\scriptsize torus}}^2 $ is
the modular parameter for the Klein bottle with $\tau=\frac{M}{N}\exp[i(\pi-\vartheta)]$.
The free energies were computed by solving a functional equation of the transfer matrices
\cite{CMP} and they were shown in \cite{CMP03}.

\begin{figure}[ht]
\setlength{\unitlength}{6.05mm}
\begin{picture}(8,14)
\put(0,0){
\begin{picture}(8,14)
\put(2,0){
\includegraphics[width=.8\linewidth]{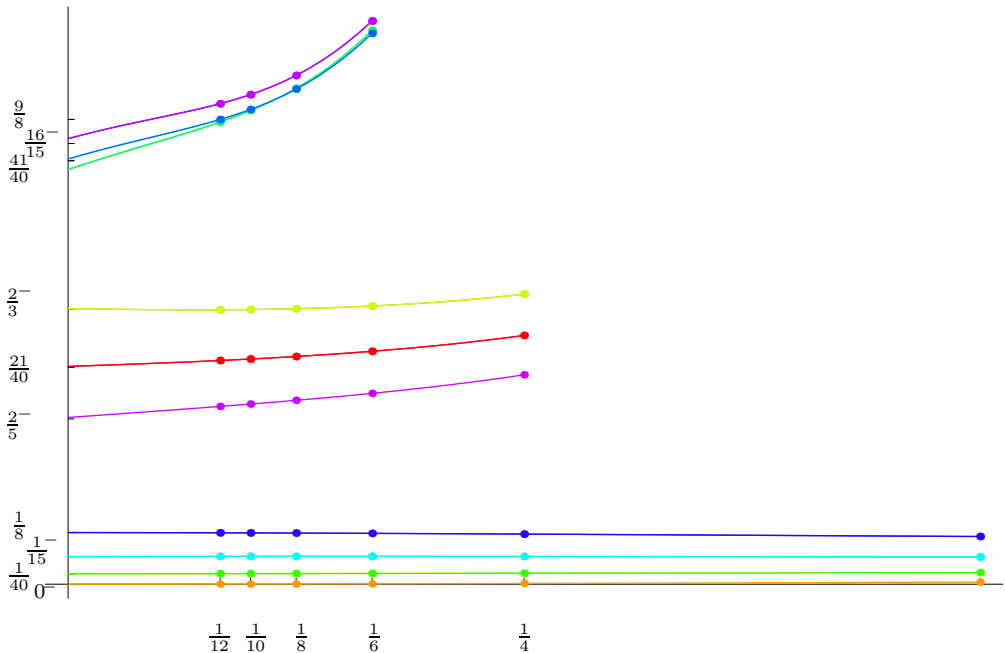}
}
\put(2.7,0.65){\small\pp{c}{0^{\!-}}}
\put(2.1,0.94){\small\pp{c}{\frac{1}{40}}}
\put(2.6,1.54){\small\pp{c}{{\frac{1}{15}}^{\!\!-}}}
\put(2.1,2.05){\small\pp{c}{\frac{1}{8}}}
\put(2.1,4.3){\small\pp{c}{{\frac{2}{5}}^{\!-}}}
\put(2.1,5.5){\small\pp{c}{\frac{21}{40}}}
\put(2.1,7){\small\pp{c}{{\frac{2}{3}}^{\!-}}}
\put(2.1,9.9){\small\pp{c}{\frac{41}{40}}}
\put(2.6,10.5){\small\pp{c}{{\frac{16}{15}}^{\!-}}}
\put(2.1,11.1){\small\pp{c}{\frac{9}{8}}}
\put(6.5,-0.4){\small\pp{c}{\frac{1}{12}}}
\put(7.3,-0.4){\small\pp{c}{\frac{1}{10}}}
\put(8.3,-0.4){\small\pp{c}{\frac{1}{8}}}
\put(9.9,-0.4){\small\pp{c}{\frac{1}{6}}}
\put(13.2,-0.4){\small\pp{c}{\frac{1}{4}}}
\end{picture} }
\end{picture}
\vspace{0.4cm}
     \caption{The extrapolated sequences of the first ten energy levels of the minimal $(A_4,A_5)$ model with
      $(1,2)$ seam on the Klein bottle for $\kappa=1,2$. The energy levels corresponding to the terms
      of negative coefficients in $K^{(2)}$ are marked with a superscript ${}^{\!-}$. The horizontal axis is $1/N$.}
\label{fig:A5-1-2}
\end{figure}

\begin{figure}[ht]
\setlength{\unitlength}{6.05mm}
\begin{picture}(8,14)
\put(0,0){
\begin{picture}(8,14)
\put(2,0){
\includegraphics[width=.8\linewidth]{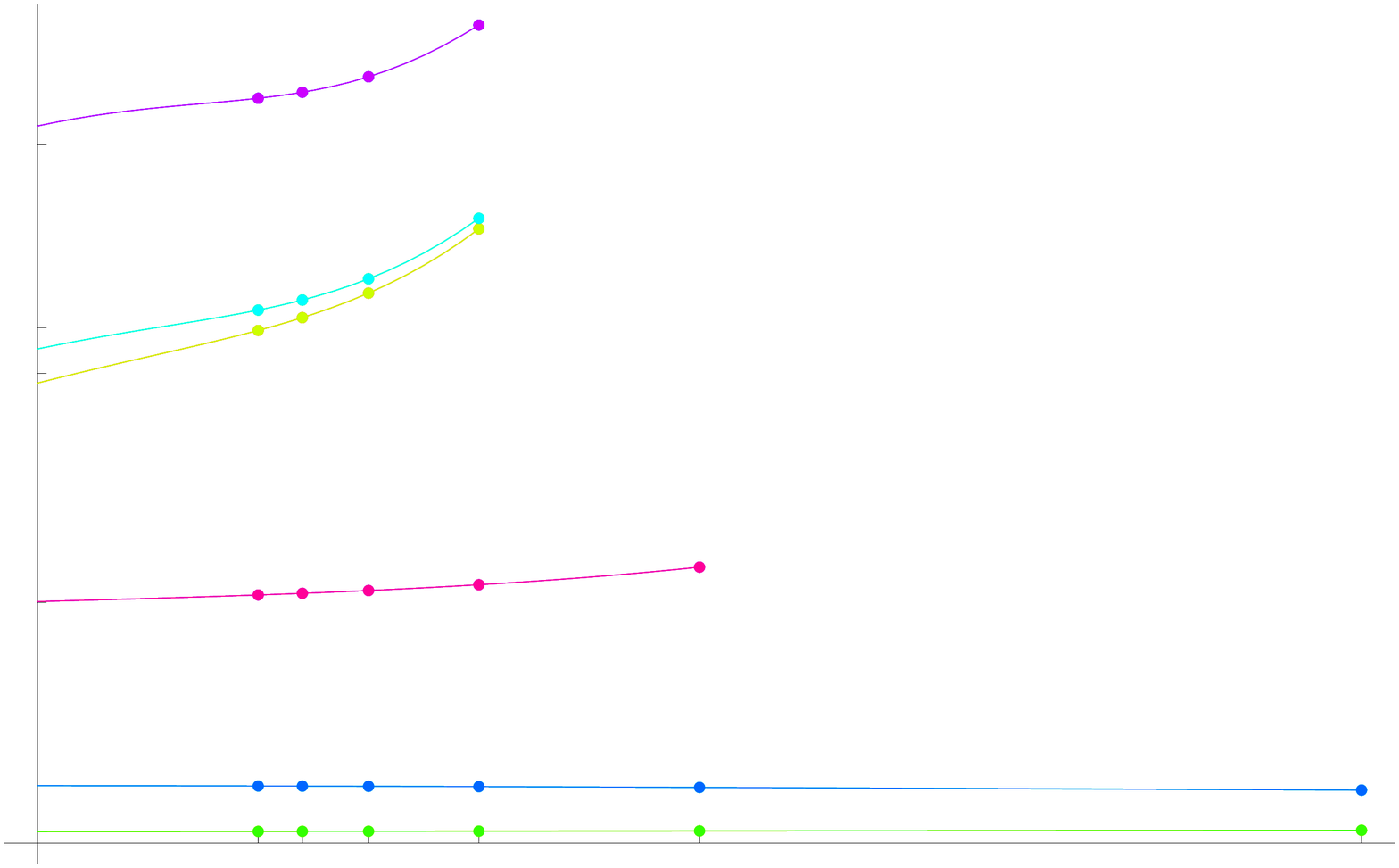}
}
\put(2.1,0.8){\small\pp{c}{\frac{1}{40}}}
\put(2.1,1.54){\small\pp{c}{{\frac{1}{8}}}}
\put(2.1,4.2){\small\pp{c}{\frac{21}{40}}}
\put(2.1,7.4){\small\pp{c}{\frac{41}{40}}}
\put(2.1,8.5){\small\pp{c}{\frac{9}{8}}}
\put(2.1,11.2){\small\pp{c}{\frac{61}{40}}}
\put(6.2,-0.4){\small\pp{c}{\frac{1}{12}}}
\put(7,-0.4){\small\pp{c}{\frac{1}{10}}}
\put(8,-0.4){\small\pp{c}{\frac{1}{8}}}
\put(9.7,-0.4){\small\pp{c}{\frac{1}{6}}}
\put(13.1,-0.4){\small\pp{c}{\frac{1}{4}}}
\end{picture} }
\end{picture}
\vspace{0.4cm}
     \caption{The extrapolated sequences of the first six energy levels of the minimal $(A_4,D_4)$ model with
      $(1,2)$ seam on the Klein bottle for $\kappa=1,2$. The only difference between the $\kappa=1,2$ cases are
      the degeneracies of the energy levels.}
\label{fig:D4-1-2}
\end{figure}

\begin{figure}[ht]
\setlength{\unitlength}{6.05mm}
\begin{picture}(8,14)
\put(0,0){
\begin{picture}(8,14)
\put(2,0){
\includegraphics[width=.8\linewidth]{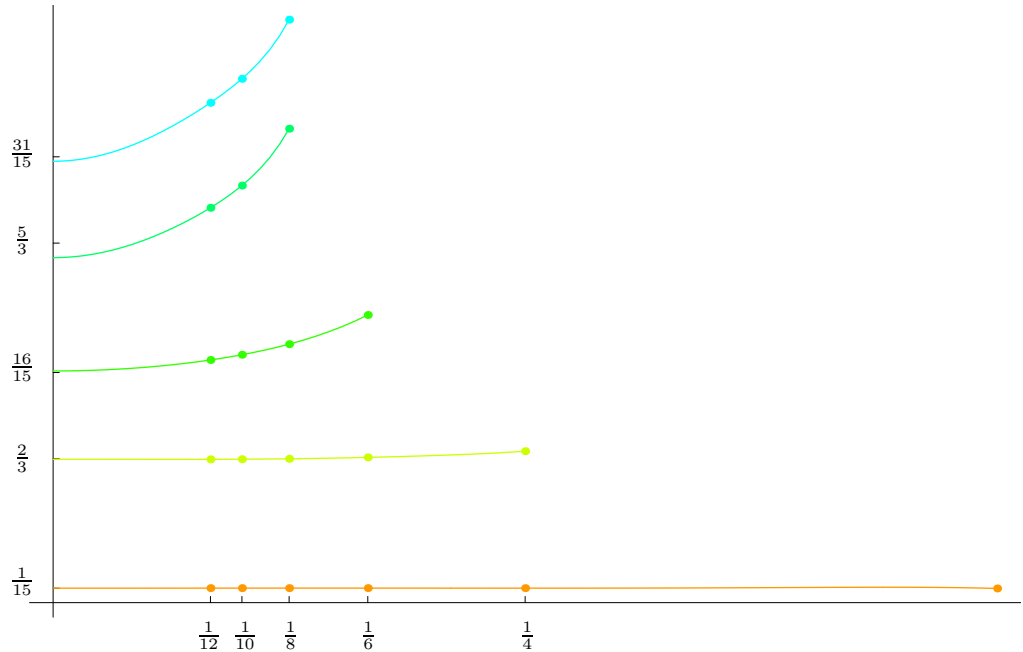}
}
\put(2.1,0.8){\small\pp{c}{\frac{1}{15}}}
\put(2.1,3.5){\small\pp{c}{\frac{2}{3}}}
\put(2.1,5.5){\small\pp{c}{\frac{16}{15}}}
\put(2.1,8.3){\small\pp{c}{\frac{5}{3}}}
\put(2.1,10.2){\small\pp{c}{\frac{31}{15}}}
\put(6.2,-0.4){\small\pp{c}{\frac{1}{12}}}
\put(7,-0.4){\small\pp{c}{\frac{1}{10}}}
\put(8,-0.4){\small\pp{c}{\frac{1}{8}}}
\put(9.7,-0.4){\small\pp{c}{\frac{1}{6}}}
\put(13.2,-0.4){\small\pp{c}{\frac{1}{4}}}
\end{picture} }
\end{picture}
\vspace{0.4cm}
     \caption{The extrapolated sequences of the first five energy levels of the minimal $(A_4,D_4)$ model with
      $(1,3)$ seam on the Klein bottle for $\kappa=2$.}
\label{fig:D4-1-3}
\end{figure}

We carried out the numerics for $A_3$, $A_4$, $A_5$, $D_4$ and $D_5$ with seams inserted 
on the Klein bottle and on the M\"obius strip. 
We fit the sets of data $\{\log T_i \,|\ N=2,4,6,\ldots\}$ onto a polynomial of the form
$a_1 N + a_0 +\frac{a_{-1}}{N}+\cdots\frac{a_{-k}}{N^k}$ and extrapolate it to $1/N\rightarrow0$.
By comparing with the expression \eqref{eq:finiteSizeCorr},
the eigenvalue spectra agree well with the result obtained from the theory of open descendants in section \ref{sec:OpenDescendants}. 
The M\"obius strip and Klein bottle partition functions of the $D_4$ and $A_5$ models are listed in tables \ref{tab:MoebiusA4D4}--\ref{tab:KleinA5}. 
In figure \ref{fig:A5-1-2} and in table \ref{fig:A5numericdata} we illustrate some typical numerical results for the $A_5$ case
with a $(1,2)$ seam inserted.
In figure \ref{fig:D4-1-2} and in table \ref{fig:D41numericdata} we show the $D_4$ case with seam $(1,2)$;
in figure \ref{fig:D4-1-3} and in  table \ref{fig:D42numericdata} we illustrate the
$D_4$ case with seam $(1,3)$ for $\kappa=2$.

\begin{table}[hbt]
{\small
\begin{equation}
\begin{array}{|c|cccccccccc|}\hline
\multicolumn{11}{|c|}{A_5:\quad (1,2)}\cr\hline
n& 0 & 1 & 2 & 3 & 4 & 5 & 6 & 7 & 8 & 9 \cr\hline
\mbox{Degeneracy}& 1&2&2&2&1&2&2&2&2&2\cr\hline
\mbox{Exact}& 0^- & \frac{1}{40} & \frac{1}{15} ^- & \frac{1}{8} & \frac{2}
   {5} ^- & \frac{21}{40} & \frac{2}{3} ^- & \frac{41}{40} & \frac{16}{15} ^- & \frac{9}
   {8} \cr\hline
   \mbox{Numerical}& 10^{-6} & 0.0250 & 0.0667 & 0.1250 & 0.4033 & 0.5260 & 0.6649 & 
   0.9779 & 0.9988 & 1.0442 \cr\hline
   |\mbox{diff.}|& 10^{-6} & 10^{-5} & 10^{-5} & 
   10^{-5} & 0.0033 & 0.0001 & 0.0018 & 0.0471 & 
 0.0679 & 0.0809\cr\hline\hline
\end{array}\nonumber
\end{equation}}
\caption{Numerical values of the exponents of the Klein bottle partition functions $K^{(\kappa)}_{(1,2,1)}$ 
for $(A_4,A_5)$ minimal model with a $(1,2)$ seam.}
\label{fig:A5numericdata}
\end{table}

%
%

\begin{table}[hbt]
{\small
\begin{equation}
\begin{array}{|c|cccccc|}\hline
\multicolumn{7}{|c|}{D_4 :\quad (1,2)}\cr\hline
n& 0 & 1 & 2 & 3 & 4 & 5 \cr\hline
\mbox{Degeneracy }(\kappa\!=\!1\, /\, 2)& 3/1&3/1&3/1&3/1&3/1&3/1\cr\hline
\mbox{Exact}& \frac{1}{40} & \frac{1}{8} & \frac{21}{40} & \frac{41}{40} & \frac{9}{8} & \frac{61}{40} \cr\hline
   \mbox{Numerical}& 0.0250 & 0.1250 & 0.5260 & 0.9779 & 1.0442 & 1.5102  \cr\hline
   |\mbox{diff.}|& 4.5\times10^{-5} & 3.3\times10^{-5} & 9.7\times10^{-4} & 0.0471 & 0.0809 & 0.0148 \cr\hline\hline
\end{array}\nonumber
\end{equation}}
\caption{Numerical values of the exponents of the Klein bottle partition functions $K^{(\kappa)}_{(1,2,1)}$ 
for $(A_4,D_4)$ minimal model with a $(1,2)$ seam. The degeneracies of the first six energy levels
for $\kappa=1$ and $2$ are three and one respectively.}
\label{fig:D41numericdata}
\end{table}

%
%
%
%

\begin{table}[hbt]
{\small
\begin{equation}
\begin{array}{|c|ccccc|}\hline
\multicolumn{6}{|c|}{D_4 :\quad \kappa=2.\  (1,3)}\cr\hline
n& 0 & 1 & 2 & 3 & 4  \cr\hline
\mbox{Exact}& \frac{1}{15} & \frac{2}{3} & \frac{16}{15} & \frac{5}{3} & \frac{31}{15} \cr\hline
   \mbox{Numerical}& 0.0671 & 0.6646 & 1.0739 & 1.5992 & 2.0640 \cr\hline
   |\mbox{diff.}|& 4.5\times10^{-4} & 0.0021 & 0.0072 & 0.0675 & 0.0207 \cr\hline\hline
\end{array}\nonumber
\end{equation}}
\caption{Numerical values of the exponents of the Klein bottle partition functions $K^{(2)}_{(1,3,1)}$ 
for $(A_4,D_4)$ minimal model with a $(1,3)$ seam.}
\label{fig:D42numericdata}
\end{table}

\section{Discussion}

In this paper we studied \ade lattice models on the Klein bottle and on the M\"obius strip
with defect lines. The non-orientable geometries were achieved by inserting a flip operator before taking the trace
of the transfer matrices. The structure of the transfer matrices is not arbitrary. They must
be consistent with the non-orientable topologies, namely, the transfer matrices have to
be left-right symmetric upon the action of the flip operator.

Furthermore, we showed that in order to realize all the open descendants of the unitary minimal models,
not only the flip on the topologies of the lattice but also a flip in the spin configuration space
must be applied. This flip in the spin configuration space is the $\Bbb Z_2$ automorphism of the graph $G$,
and essentially it corresponds to the simple current of the Ocneanu fusion algebra.
We diagonalized the finite-size lattice transfer matrices numerically and  verified that the extrapolated partition functions
agree well with the prediction of the theory of open descendants.

Finally, we remark that the methods of this paper can be applied to $\Bbb Z_k$ parafermionic models and other rational conformal field theories whose lattice realizations are known. 

\section*{Acknowledgements} CHOC thanks IPAM, UCLA for their hospitality in the \textit{2nd Reunion Conference on CFT} 
held at Lake Arrowhead during which part of the work were done. 
PAP acknowledges support by the Australian Research Council.

\begin{table}[b]
\small
  \centering
  \begin{eqnarray*}
M_{(1,1)}^{(1)} &=&     \chi _{1,1}+\chi _{1,5} \\
M_{(1,2)}^{(1)} &=& \chi _{1,1}+2 \chi _{1,3}-\chi _{1,5} \\
M_{(1,3)}^{(1)} &=& \chi _{1,1}+\chi _{1,5}\quad = M_{(1,4)}^{(1)} \\
M_{(3,1)}^{(1)} &=& \chi _{1,1}+\chi _{1,5}+\chi _{3,1}+\chi _{3,5}\\
M_{(3,2)}^{(1)} &=& \chi _{1,1}+2 \chi _{1,3}-\chi _{1,5}+\chi _{3,1}+2 \chi_{3,3}-\chi_{3,5} \\
M_{(3,3)}^{(1)} &=& \chi _{1,1}+\chi _{1,5}+\chi _{3,1}+\chi _{3,5} \quad = M_{(3,4)}^{(1)}\\
M_{(1,1)}^{(2)} &=& \chi _{1,1}-\chi _{1,5} \\
M_{(1,2)}^{(2)} &=& \chi _{1,1}+\chi _{1,5} \\
M_{(1,3)}^{(2)} &=& \chi _{1,3}\quad =M_{(1,4)}^{(2)} \\
M_{(3,1)}^{(2)} &=& \chi _{1,1}-\chi _{1,5}+\chi _{3,1}-\chi _{3,5} \\
M_{(3,2)}^{(2)} &=&\chi _{1,1}+\chi _{1,5}+\chi _{3,1}+\chi _{3,5} \\
M_{(3,3)}^{(2)} &=&\chi _{1,3}+\chi _{3,3}\quad = M_{(3,4)}^{(2)}
  \end{eqnarray*}
  \caption{The $(A_4,D_4)$ M\"obius strip partition functions of the 3-state Potts model.}
  \label{tab:MoebiusA4D4}
\end{table}

\begin{table}[p]
\small
  \centering
  \begin{eqnarray*}
K_{(1,1)}^{(1)} &=&\chi _{1,1}+2 \chi _{1,3}+\chi _{1,5}+\chi _{3,1}+2 \chi _{3,3}+\chi
_{3,5} \\
K_{(1,2)}^{(1)} &=&3 \chi _{1,2}+3 \chi _{1,4}+3 \chi _{3,2}+3 \chi _{3,4}\\
K_{(1,3)}^{(1)} &=&\chi _{1,1}+2 \chi _{1,3}+\chi _{1,5}+\chi _{3,1}+2 \chi _{3,3}+\chi
_{3,5} \quad = K_{(1,4)}^{(1)} \\
K_{(3,1)}^{(1)} &=&\chi _{1,1}+2 \chi _{1,3}+\chi _{1,5}+2 \chi _{3,1}+4 \chi _{3,3}+2
\chi _{3,5}\\
K_{(3,2)}^{(1)} &=&3 \chi _{1,2}+3 \chi _{1,4}+6 \chi _{3,2}+6 \chi _{3,4} \\
K_{(3,3)}^{(1)} &=&\chi _{1,1}+2 \chi _{1,3}+\chi _{1,5}+2 \chi _{3,1}+4 \chi _{3,3}+2
\chi _{3,5} \quad = K_{(3,4)}^{(1)} \\
K_{(1,1)}^{(2)} &=& \chi _{1,1}+\chi _{1,5}+\chi _{3,1}+\chi _{3,5}\\
K_{(1,2)}^{(2)} &=& \chi _{1,2}+\chi _{1,4}+\chi _{3,2}+\chi _{3,4}\\
K_{(1,3)}^{(2)} &=& \chi _{1,3}+\chi _{3,3} \quad = K_{(1,4)}^{(2)} \\
K_{(3,1)}^{(2)} &=& \chi _{1,1}+\chi _{1,5}+2 \chi _{3,1}+2 \chi _{3,5}\\
K_{(3,2)}^{(2)} &=& \chi _{1,2}+\chi _{1,4}+2 \chi _{3,2}+2 \chi _{3,4}\\
K_{(3,3)}^{(2)} &=& \chi _{1,3}+2 \chi _{3,3} \quad = K_{(3,4)}^{(2)} 
  \end{eqnarray*}
  \caption{The $(A_4,D_4)$ Klein bottle partition functions of the 3-state Potts model.}
  \label{tab:KleinA4D4}
\end{table}

\begin{table}[p]
\small
  \centering
  \begin{eqnarray*}
M_{(1,1)}^{(1)} &=& \chi _{1,1}\\
M_{(1,2)}^{(1)} &=& \chi _{1,1}+\chi _{1,3}\\
M_{(1,3)}^{(1)} &=& \chi _{1,1}+\chi _{1,3}+\chi _{1,5}\\
M_{(3,1)}^{(1)} &=& \chi _{1,1}+\chi _{3,1}\\
M_{(3,2)}^{(1)} &=& \chi _{1,1}+\chi _{1,3}+\chi _{3,1}+\chi _{3,3}\\
M_{(3,3)}^{(1)} &=& \chi _{1,1}+\chi _{1,3}+\chi _{1,5}+\chi _{3,1}+\chi _{3,3}+\chi
_{3,5}\\
M_{(1,1)}^{(2)} &=& \chi _{1,5}\\
M_{(1,2)}^{(2)} &=& \chi _{1,3}-\chi _{1,5}\\
M_{(1,3)}^{(2)} &=& \chi _{1,1}-\chi _{1,3}+\chi _{1,5}\\
M_{(3,1)}^{(2)} &=& \chi _{1,5}+\chi _{3,5}\\
M_{(3,2)}^{(2)} &=& \chi _{1,3}-\chi _{1,5}+\chi _{3,3}-\chi _{3,5}\\
M_{(3,3)}^{(2)} &=& \chi _{1,1}-\chi _{1,3}+\chi _{1,5}+\chi _{3,1}-\chi _{3,3}+\chi
_{3,5} %
  \end{eqnarray*}
  \caption{The $(A_4,A_5)$ M\"obius strip partition functions.
  They satisfy \eqref{eq:unitaryboundarysymmetry} and $M_{(r,s)}^{(\kappa)}=M_{(r,6-s)}^{(\kappa)} $.}
  \label{tab:MoebiusA5}
\end{table}

\begin{table}[t]
\small
  \centering
  \begin{eqnarray*}
K_{(1,1)}^{(1)} &=& \chi _{1,1}+\chi _{1,2}+\chi _{1,3}+\chi _{1,4}+\chi _{1,5}+\chi
_{3,1}+\chi _{3,2}+\chi _{3,3}+\chi _{3,4}+\chi _{3,5}\\
K_{(1,2)}^{(1)} &=& \chi _{1,1}+2 \chi _{1,2}+2 \chi _{1,3}+2 \chi _{1,4}+\chi _{1,5}+\chi
_{3,1}+2 \chi _{3,2}+2 \chi _{3,3}+2 \chi _{3,4}+\chi _{3,5}\\
K_{(1,3)}^{(1)} &=& \chi _{1,1}+2 \chi _{1,2}+3 \chi _{1,3}+2 \chi _{1,4}+\chi _{1,5}+\chi
_{3,1}+2 \chi _{3,2}+3 \chi _{3,3}+2 \chi _{3,4}+\chi _{3,5}\\
K_{(3,1)}^{(1)} &=& \chi _{1,1}+\chi _{1,2}+\chi _{1,3}+\chi _{1,4}+\chi _{1,5}+2 \chi
_{3,1}+2 \chi _{3,2}+2 \chi _{3,3}+2 \chi _{3,4}+2 \chi _{3,5}\\
K_{(3,2)}^{(1)} &=& \chi _{1,1}+2 \chi _{1,2}+2 \chi _{1,3}+2 \chi _{1,4}+\chi _{1,5}+2
\chi _{3,1}+4 \chi _{3,2}+4 \chi _{3,3}+4 \chi _{3,4}+2 \chi _{3,5}\\
K_{(3,3)}^{(1)} &=& \chi _{1,1}+2 \chi _{1,2}+3 \chi _{1,3}+2 \chi _{1,4}+\chi _{1,5}+2
\chi _{3,1}+4 \chi _{3,2}+6 \chi _{3,3}+4 \chi _{3,4}+2 \chi _{3,5}\\
K_{(1,1)}^{(2)} &=& \chi _{1,1}-\chi _{1,2}+\chi _{1,3}-\chi _{1,4}+\chi _{1,5}+\chi
_{3,1}-\chi _{3,2}+\chi _{3,3}-\chi _{3,4}+\chi _{3,5}\\
K_{(1,2)}^{(2)} &=& -\chi _{1,1}+2 \chi _{1,2}-2 \chi _{1,3}+2 \chi _{1,4}-\chi _{1,5}-\chi
_{3,1}+2 \chi _{3,2}-2 \chi _{3,3}+2 \chi _{3,4}-\chi _{3,5}\\
K_{(1,3)}^{(2)} &=& \chi _{1,1}-2 \chi _{1,2}+3 \chi _{1,3}-2 \chi _{1,4}+\chi _{1,5}+\chi
_{3,1}-2 \chi _{3,2}+3 \chi _{3,3}-2 \chi _{3,4}+\chi _{3,5}\\
K_{(3,1)}^{(2)} &=& \chi _{1,1}-\chi _{1,2}+\chi _{1,3}-\chi _{1,4}+\chi _{1,5}+2 \chi
_{3,1}-2 \chi _{3,2}+2 \chi _{3,3}-2 \chi _{3,4}+2 \chi _{3,5}\\
K_{(3,2)}^{(2)} &=& -\chi _{1,1}+2 \chi _{1,2}-2 \chi _{1,3}+2 \chi _{1,4}-\chi _{1,5}-2
\chi _{3,1}+4 \chi _{3,2}-4 \chi _{3,3}+4 \chi _{3,4}-2 \chi _{3,5}\\
K_{(3,3)}^{(2)} &=& \chi _{1,1}-2 \chi _{1,2}+3 \chi _{1,3}-2 \chi _{1,4}+\chi _{1,5}+2
\chi _{3,1}-4 \chi _{3,2}+6 \chi _{3,3}-4 \chi _{3,4}+2 \chi _{3,5}
  \end{eqnarray*}
  \caption{The $(A_4,A_5)$ Klein bottle partition functions.
  The twisted boundary conditions satisfy \eqref{eq:unitaryboundarysymmetry} and $K_{(r,s)}^{(\kappa)}=K_{(r,6-s)}^{(\kappa)} $.}
  \label{tab:KleinA5}
\end{table}

\clearpage

\bibliographystyle{unsrt}
\bibliography{reference}
\end{document}